\documentclass[journal,11pt,onecolumn]{IEEEtran}
\usepackage{amsmath,amssymb}
\usepackage{amsthm}
\usepackage{txfonts}
\usepackage{bm}
\usepackage{graphicx}
\usepackage{xcolor}
\usepackage{tikz}
\usetikzlibrary{arrows.meta}
\usepackage{cite}
\usepackage{microtype}
\usepackage[colorlinks,linkcolor=blue,citecolor=blue,urlcolor=blue]{hyperref}
\hypersetup{pdftitle={Optimal (1,1)-Criss-Cross Deletion-Correcting Codes with Linear-Time Encoder and Decoder},
            pdfauthor={Zhiqiang Li, Yubo Sun, and Gennian Ge}}
\newtheorem{theorem}{Theorem}
\newtheorem{lemma}{Lemma}
\newtheorem{definition}{Definition}
\newtheorem{construction}{Construction}
\newtheorem{corollary}{Corollary}
\newtheorem{remark}{Remark}
\newtheorem{example}{Example}

\DeclareMathOperator{\Rank}{Rank}
\DeclareMathOperator{\Syn}{Syn}
\DeclareMathOperator{\VTSyn}{VTSyn}
\newcommand{\Dball}{\mathbb D}
\newcommand{\E}{\mathbb E}
\newcommand{\ind}{\mathbf{1}}
\definecolor{seedcolor}{RGB}{202,224,198}
\definecolor{innercolor}{RGB}{201,219,237}
\definecolor{outercolor}{RGB}{248,223,191}
\definecolor{paritycolor}{RGB}{222,208,235}
\allowdisplaybreaks[1]
\title{Optimal \texorpdfstring{$(1,1)$}{(1,1)}-Criss-Cross Deletion-Correcting Codes with Linear-Time Encoder and Decoder}

\begin{document}

\author{Zhiqiang~Li, Yubo~Sun, and Gennian~Ge%
\thanks{This work was supported in part by the National Key Research and Development Program of China under Grant 2025YFC3409900, in part by the National Natural Science Foundation of China under Grant 12231014, in part by Beijing Scholars Program, and in part by the Postdoctoral Fellowship Program and China Postdoctoral Science Foundation under Grant Number BX2026006. \emph{(Corresponding author: Yubo Sun.)}}
\thanks{Zhiqiang Li ({\tt lzq88@mail.ustc.edu.cn}) and Gennian Ge ({\tt gnge@zju.edu.cn}) are with the School of Mathematical Sciences, Capital Normal University, Beijing 100048, China.}%
\thanks{Yubo Sun ({\tt ybsun@cnu.edu.cn}) is with the Institute of Mathematics and Interdisciplinary Sciences, Xidian University, Xi'an 710126, China.}}

\maketitle

\begin{abstract}
We consider a $q$-ary $n\times n$ array subject to a $(1,1)$-criss-cross deletion, which removes one row and one column. We first derive a sphere-packing type lower bound on the redundancy of codes correcting such deletions. For every $q\geq 2$ and $n\geq 27$, the redundancy is greater than $2n+2\log_q n-2$. For fixed $q$, it is at least $2n-1+2\log_q n-o_q(1)$ as $n\to\infty$. We then present two code constructions. The first construction achieves redundancy $2n-1+2\log_q n+o_q(1)$ by selecting a largest syndrome class. It matches the lower bound up to a vanishing term and admits $O_q(n^2)$-time decoding. The second construction has redundancy $2n+2\log_q n+O_q(1)$ and admits $O_q(n^2)$-time algorithms for both encoding and decoding. For fixed $q$, these running times are linear in the number of array entries.
\end{abstract}


\section{Introduction}\label{sec:introduction}
\IEEEPARstart{T}{wo-dimensional} error-correcting codes represent codewords as arrays. They arise in applications such as QR codes~\cite{hagiwara2023multi}, DNA-based storage~\cite{heckel2019characterization}, and racetrack memories~\cite{Chee2018Coding,Chee2018Codes}. A substantial part of the literature deals with substitutions and erasures in arrays~\cite{Yaakobi2010High,Abdel1988Two,Blaum2000MDS,gabidulin1985optimum,gabidulin2008error,roth1991maximum,roth1997probabilistic,sidorenko1976class,wachter2017list}. After a row or column is deleted, the decoder must recover the array from the surviving entries with the deletion positions unknown. This synchronization problem has been studied for several row and column error models~\cite{wang2026binary,bitar2021,chee2021,welter2022,sun2026,zhang2026almost,hagiwara2020conversion}.

Bitar et al.~\cite{bitar2021} introduced criss-cross deletions. A $(t_r,t_c)$-criss-cross deletion removes $t_r$ rows and $t_c$ columns from an $n\times n$ array, where $1\leq t_r,t_c\leq n$. Let $\Sigma_q=\{0,1,\ldots,q-1\}$. A code $\mathcal C\subseteq\Sigma_q^{n\times n}$ is a $(t_r,t_c)$-criss-cross deletion-correcting code if every codeword can be recovered from the resulting $(n-t_r)\times(n-t_c)$ array. Its redundancy, measured in $q$-ary symbols, is $r_{\mathcal C}=n^2-\log_q|\mathcal C|$. In this paper, we focus on the case $(t_r,t_c)=(1,1)$.

For $(1,1)$-criss-cross deletion-correcting codes, Bitar et al.~\cite{bitar2021} established a lower bound of $2n+2\log_q n-3$ for $n\geq 41$ and an asymptotic lower bound of $2n+2\log_q n-2$. They also constructed codes with redundancy $2n+4\log_q n+O_q(1)$. For the binary alphabet, Chee et al.~\cite{chee2021} reduced the redundancy to $2n+2\log_2 n+O(\log_2\log_2 n)$. Sun and Ge~\cite{sun2026} constructed codes for arbitrary alphabets with redundancy $2n+2\log_q n+O_q(1)$ and gave efficient decoders. The sizes of those codes were obtained by choosing large syndrome classes, and constructing efficient encoders at the same redundancy was left open.

Few known constructions admit efficient encoding, and those that do generally require higher redundancy. For binary arrays with $n$ a power of two, Bitar et al.~\cite{bitar2021} presented an explicit encoder with redundancy at most $2n+9\log_2 n+O(1)$. For $q\geq 3$ and $n\geq 11$, Liu et al.~\cite{liu2026criss} gave an efficient encoder with redundancy at most $2n+2\log_q n+(2n-13)\log_q\frac{q}{q-1}+12$.
The term $(2n-13)\log_q\frac{q}{q-1}$ grows linearly with $n$ for fixed $q$ and is bounded when $q=\Omega(n)$.

In this paper, we first improve the lower bound for every $q\geq 2$. The main observation is that if two distinct deletion patterns applied to the same codeword give the same output, then its deletion ball contains an array in which two adjacent rows or two adjacent columns have a common single-deletion output. The number of arrays with such a pair is small. A sphere-packing argument based on this observation gives $r_{\mathcal C}>2n+2\log_q n-2$ for $n\geq 27$ and the asymptotic lower bound $2n-1+2\log_q n-o_q(1)$ for every fixed $q$.

We then present two constructions for fixed $q$. In the first construction, we use the row and column parity checks to recover the missing entries. To restore the row position, we group the columns by their numbers of zeros and record the number of zeros of each row in each group. This representation is unchanged by a column permutation and distinguishes adjacent rows with high probability, including in the binary case. We apply a Varshamov--Tenengolts (VT) constraint to the row representations and a second VT constraint to the full columns. The resulting codes have redundancy $2n-1+2\log_q n+o_q(1)$, which is optimal up to a vanishing term.

In the second construction, the upper-left $v\times v$ block $\bm A$ determines canonical row and column orders. These orders let us read the column representation before restoring the row position and the row representation before restoring the column position. In each direction we impose the same VT syndrome on two sequences. After parity completion and canonical reordering, the deletion can change the order in at most one of the two sequences in each direction, so the other supplies the syndrome for decoding. To encode a prescribed syndrome, we reserve one column vector and choose its insertion position after the other columns have been filled. This gives redundancy $2n+2\log_q n+O_q(1)$. Both decoders recover the original array in $O_q(n^2)$ time. The second decoder also determines both deletion positions. Encoding and recovering the message from the corrected array in the second construction each take $O_q(n^2)$ time.

The remainder of this paper is organized as follows. Section~\ref{sec:preliminaries} introduces the notation, error model, and tools used throughout the paper. Section~\ref{sec:lower} establishes the lower bound. Sections~\ref{sec:dense} and~\ref{sec:efficient} present the two constructions and their decoders. Section~\ref{sec:efficient} also gives an encoder. Section~\ref{sec:conclusion} concludes the paper and discusses a direction for further work.

\section{Preliminaries}\label{sec:preliminaries}

\subsection{Notations}
For integers $i$ and $j$, let $[i:j]=\{i,i+1,\ldots,j\}$, with $[i:j]=\varnothing$ if $i>j$. For a vector $\bm x=(x_1,\ldots,x_m)$, write $\bm x_{[i:j]}=(x_i,\ldots,x_j)$. For an array $\bm X=(X_{i,j})\in\Sigma_q^{m\times n}$, denote by $\bm X_{i,[1:n]}$ and $\bm X_{[1:m],j}$ its $i$th row and $j$th column, respectively. The subarray on rows $[i_1:i_2]$ and columns $[j_1:j_2]$ is denoted by $\bm X_{[i_1:i_2],[j_1:j_2]}$. Write $\bm X^{\mathrm T}$ for its transpose.

Two distinct vectors of the same length are ordered lexicographically by comparing their first unequal coordinates. The rank of $\bm x\in\Sigma_q^m$ in this order is $\Rank(\bm x)=\sum_{i=1}^{m}x_i q^{m-i}\in[0:q^m-1]$. We use the same notation with the appropriate alphabet for vectors over $\Sigma_{n+1}$. For a rectangular array, write $\Rank^r(\bm X)$ and $\Rank^c(\bm X)$ for the sequences of ranks of its full rows and columns, respectively.

Unless stated otherwise, asymptotic statements are for fixed $q$. The notation $O_q(\cdot)$ allows the implied constant to depend on $q$, and $o_q(1)$ denotes a term tending to zero as $n\to\infty$ for fixed $q$.

\subsection{Error Models and Codes}
For $\bm X\in\Sigma_q^{n\times n}$, let $\Delta_{i,j}(\bm X)$ denote the array obtained by deleting row $i$ and column $j$. Explicitly,
\[
\Delta_{i,j}(\bm X)=
\begin{bmatrix}
\bm X_{[1:i-1],[1:j-1]} & \bm X_{[1:i-1],[j+1:n]}\\
\bm X_{[i+1:n],[1:j-1]} & \bm X_{[i+1:n],[j+1:n]}
\end{bmatrix}.
\]
The set $\Dball_{1,1}(\bm X)=\{\Delta_{i,j}(\bm X):i,j\in[1:n]\}$ is called the $(1,1)$-criss-cross deletion ball of $\bm X$. For a vector $\bm x$ of length $m$, let $\Delta_i(\bm x)$ denote the result of deleting its $i$th coordinate. Its single-deletion ball is $\Dball_1(\bm x)=\{\Delta_i(\bm x):i\in[1:m]\}$. A single insertion is the reverse operation.

\begin{definition}\label{def:code}
Let $q,n\geq 2$. A nonempty set $\mathcal C\subseteq\Sigma_q^{n\times n}$ is a $q$-ary $(1,1)$-criss-cross deletion-correcting code if $\Dball_{1,1}(\bm X)\cap\Dball_{1,1}(\bm X')=\varnothing$ for distinct $\bm X,\bm X'\in\mathcal C$. Its redundancy is $r_{\mathcal C}=n^2-\log_q|\mathcal C|$.
\end{definition}

A nonempty set of length-$m$ vectors is a single deletion-correcting code if its single-deletion balls are pairwise disjoint. For a $q$-ary code $\mathcal C$ of this type, the redundancy is $m-\log_q|\mathcal C|$. The corresponding array insertion model adds one row and one column. The following equivalence from~\cite{bitar2021} allows us to study deletions. See also~\cite{welter2022} for multiple row and column errors.

\begin{lemma}[{\cite[Theorem 1]{bitar2021}}]\label{lem:ins-del}
A code $\mathcal C\subseteq\Sigma_q^{n\times n}$ corrects a $(1,1)$-criss-cross deletion if and only if it corrects a $(1,1)$-criss-cross insertion.
\end{lemma}

\subsection{Useful Tools}
We first recall the VT syndrome for single-deletion correction. For a sequence $\bm x=(x_1,\ldots,x_m)$ over an ordered alphabet, its signature $\alpha(\bm x)\in\Sigma_2^{m-1}$ has entries $\alpha(\bm x)_i=\ind_{\{x_{i+1}\geq x_i\}}$. Let $\Syn(\bm x)=\sum_{i=1}^m i x_i$ be the VT syndrome of $\bm{x}$. For $m\geq 2$, we set
\[
\VTSyn(\bm x)=\Syn(\alpha(\bm x))\pmod m,
\]
with the residue taken in $[0:m-1]$. The $q$-ary VT class $\mathrm{VT}_{q,m}(\beta,\gamma)$ consists of the vectors $\bm x\in\Sigma_q^m$ satisfying $\sum_{i=1}^m x_i\equiv\beta\pmod q$ and $\VTSyn(\bm x)=\gamma$.

\begin{lemma}[{\cite[Theorem 1]{tenengolts1984}}]\label{thm:vt}
For $q,m\geq 2$, $\beta\in\Sigma_q$, and $\gamma\in\Sigma_m$, every nonempty class $\mathrm{VT}_{q,m}(\beta,\gamma)$ is a single-deletion-correcting code with an $O_q(m)$-time decoder.
\end{lemma}

\begin{corollary}[{\cite[Corollary 1]{sun2026}}]\label{cor:known-symbol}
Let $\bm y$ be obtained from a sequence $\bm x$ of length $m\geq 2$ by deleting a known symbol $\sigma$. Given $\VTSyn(\bm x)$,  one can recover $\bm{x}$ in $O(m)$-time. If adjacent symbols of $\bm x$ are distinct, the deletion position can also be determined.
\end{corollary}

Here the symbols can themselves be vectors. We can compare them in lexicographic order without forming their ranks. We will also use the following probability bounds.

\begin{lemma}[{\cite[Lemma B.1]{giraud2021}}]\label{lem:markov}
Let $Z$ be a real random variable and $a\in\mathbb R$. Then $\Pr(Z\geq a)\leq e^{-a}\E[e^Z]$ and $\Pr(Z\leq a)\leq e^a\E[e^{-Z}]$ whenever the respective expectations are finite.
\end{lemma}

\begin{lemma}[{\cite[Equation (2.4)]{baillon2016}}]\label{lem:bin}
Let $B\sim\mathrm{B}(m,p)$ denote a binomial random variable, i.e., the number of successes in $m$ independent trials, each succeeding with probability $p$, with $m\geq 1$ and $0<p<1$. Then $\max_i\Pr(B=i)\leq 1/\sqrt{2emp(1-p)}$. In particular, if $p=1/q$ with $q\geq 2$, then $\max_i\Pr(B=i)\leq q/\sqrt{m+1}$.
\end{lemma}

\section{A Sphere-Packing Type Lower Bound}\label{sec:lower}
In this section, we bound the size of a $(1,1)$-criss-cross deletion-correcting code by a sphere-packing argument.
An $n\times n$ array has $n^2$ pairs of row and column deletion
positions. If two such pairs give the same output, then the deletion
ball contains an array with the property defined below.

\begin{definition}\label{def:bad}
An $(n-1)\times(n-1)$ array $\bm Y$ is called bad if two adjacent rows of $\bm Y$ have intersecting single-deletion balls, or if two adjacent columns have intersecting single-deletion balls. All other such arrays are called good.
\end{definition}

\begin{lemma}\label{lem:badarray}
Let $q\geq 2$ and $n\geq 3$. If $\bm X\in\Sigma_q^{n\times n}$ satisfies $|\Dball_{1,1}(\bm X)|<n^2$, then $\Dball_{1,1}(\bm X)$ contains a bad array.
\end{lemma}
\begin{IEEEproof}
Since $|\Dball_{1,1}(\bm X)|<n^2$, there are distinct deletion patterns $(i_1,j_1)$ and $(i_2,j_2)$ with equal outputs. Suppose first that $i_1=i_2=i$, and assume $j_1<j_2$. After deleting row $i$, deleting either column $j_1$ or column $j_2$ gives the same sequence of columns. Thus, columns $j_1,\ldots,j_2$ in that array are equal. Choose $k\notin\{j_1,j_1+1\}$. Then $\Delta_{i,k}(\bm X)$ retains two equal adjacent columns and is bad. If $j_1=j_2$, the same argument applies to $\bm X^{\mathrm T}$.

It remains to consider $i_1\neq i_2$ and $j_1\neq j_2$. Assume $i_1<i_2$. Comparing row $i_1$ of the two equal outputs gives
\[
\Delta_{j_1}(\bm X_{i_1+1,[1:n]})
=\Delta_{j_2}(\bm X_{i_1,[1:n]}).
\]
Choose $k\notin\{i_1,i_1+1\}$ and consider $\Delta_{k,j_1}(\bm X)$. The two rows of $\bm{X}$ originally at positions $i_1$ and $i_1+1$ remain adjacent in $\Delta_{k,j_1}(\bm X)$ after coordinate $j_1$ is deleted. By the displayed equality these rows are $\Delta_{j_1}(\bm X_{i_1,[1:n]})$ and $\Delta_{j_2}(\bm X_{i_1,[1:n]})$. Both rows have a common single-deletion output, namely the subsequence
of $\bm X_{i_1,[1:n]}$ obtained by deleting its symbols at positions
$j_1$ and $j_2$. Thus, $\Delta_{k,j_1}(\bm X)$ is bad.
\end{IEEEproof}

\begin{lemma}\label{lem:countbad}
Let $\mathcal B_{n-1}$ denote the set of bad arrays. For $q\geq 2$ and $n\geq 3$, we have
\[
|\mathcal B_{n-1}|
\leq 2(n-2)q^{(n-1)(n-2)}q(n-1)^2.
\]
\end{lemma}
\begin{IEEEproof}
Fix $i\in[1:n-2]$ and consider arrays in which rows $i$ and $i+1$
have a common single-deletion output. There are $q^{(n-1)(n-2)}$
ways to choose the rows other than row $i+1$. Once row $i$ is fixed,
row $i+1$ can be obtained by deleting one symbol from row $i$ and
then inserting one symbol. The two positions and the inserted symbol
give at most $q(n-1)^2$ choices for row $i+1$.

The same bound holds for any fixed pair of adjacent columns by
transposition. Summing over the $n-2$ adjacent row pairs and the
$n-2$ adjacent column pairs gives the result.
\end{IEEEproof}

\begin{theorem}\label{thm:lower}
Let $q\geq 2$ and $n\geq 3$, and let
$\mathcal C\subseteq\Sigma_q^{n\times n}$ be a
$(1,1)$-criss-cross deletion-correcting code. Set
$\epsilon_{q,n}=2(n-2)(n-1)^3(n+1)/q^{n-2}$. Then
\begin{equation}\label{eq:redundancy-lower}
|\mathcal C|\leq \frac{q^{(n-1)^2}}{n^2}(1+\epsilon_{q,n})
\qquad\text{and}\qquad
r_{\mathcal C}\geq 2n-1+2\log_q n-\log_q(1+\epsilon_{q,n}).
\end{equation}
\end{theorem}
\begin{IEEEproof}
We partition $\mathcal C$ into two disjoint subsets. Set
$\mathcal C_1=\{\bm X\in\mathcal C:|\Dball_{1,1}(\bm X)|=n^2\}$
and $\mathcal C_2=\mathcal C\setminus\mathcal C_1$.
Since $\mathcal C$ is a $(1,1)$-criss-cross deletion-correcting code,
the deletion balls of its codewords are pairwise disjoint. Hence
\[
q^{(n-1)^2}
\geq \left|\bigcup_{\bm X\in\mathcal C}\Dball_{1,1}(\bm X)\right|
= \sum_{\bm X\in\mathcal C}|\Dball_{1,1}(\bm X)|
\geq n^2|\mathcal C_1|+|\mathcal C_2|.
\]

For each $\bm X\in\mathcal C_2$, Lemma~\ref{lem:badarray} gives
a bad array in $\Dball_{1,1}(\bm X)$. The arrays obtained in this
way are distinct since the deletion balls are disjoint. Therefore,
$|\mathcal C_2|\leq|\mathcal B_{n-1}|$. Combining these bounds gives
\begin{align*}
|\mathcal C|
&=|\mathcal C_1|+|\mathcal C_2|
\leq \frac{q^{(n-1)^2}}{n^2}+\left(1-\frac{1}{n^2}\right)|\mathcal B_{n-1}|
\leq \frac{q^{(n-1)^2}}{n^2}(1+\epsilon_{q,n}),
\end{align*}
where the last inequality follows from Lemma~\ref{lem:countbad}.
The redundancy bound follows by taking logarithms and using $r_{\mathcal C}=n^2-\log_q|\mathcal C|$.
\end{IEEEproof}

\begin{corollary}\label{cor:lowerbound}
For $q\geq 2$ and $n\geq 27$, a $(1,1)$-criss-cross deletion-correcting code $\mathcal C$ has $r_{\mathcal C}>2n+2\log_q n-2$. For every fixed $q$, it also satisfies $r_{\mathcal C}\geq 2n-1+2\log_q n-o_q(1)$ as $n\to\infty$.
\end{corollary}
\begin{IEEEproof}
For $n\geq 27$, we have
\[
\frac{\epsilon_{q,n+1}}{\epsilon_{q,n}}
=\frac{1}{q}\cdot\frac{n-1}{n-2}\left(\frac{n}{n-1}\right)^3\frac{n+2}{n+1}<1.
\]
Indeed, the three factors other than $1/q$ decrease with $n$, and their product at $n=27$ is less than $2$. Therefore
\[
\epsilon_{q,n}\leq\epsilon_{2,27}=\frac{2\cdot25\cdot26^3\cdot28}{2^{25}}<1.
\]
Since $q\geq 2$, we have $\log_q(1+\epsilon_{q,n})<1$. The finite-length bound follows from Theorem~\ref{thm:lower}. For fixed $q$, $\epsilon_{q,n}$ tends to zero.
\end{IEEEproof}

The finite-length bound improves the bound $2n+2\log_q n-3$ for $n\geq 41$ in~\cite{bitar2021}. The asymptotic bound improves the constant term in their lower bound $2n+2\log_q n-2$.

\section{Construction of Codes with Matching Redundancy}\label{sec:dense}

The construction in this section meets the lower bound in Corollary~\ref{cor:lowerbound} up to a term tending to zero. Let $\bm Y$ be obtained by deleting row $i^*$ and column $j^*$ from $\bm X$. As in~\cite[Section V]{sun2026}, we use zero row and column sums to recover the deleted entries. The negative of each surviving row sum modulo $q$ gives its missing entry. We append these entries as the last column. The column sums then give the deleted row, including the entry at $(i^*,j^*)$. Appending that row yields $\bm X'$ with row $i^*$ moved to the bottom and column $j^*$ to the right.

We restore the row position first. At this point, the full vector of a surviving row may be read in a different column order. To apply a single-deletion decoder to the sequence of rows, we need a representation of each row that is unchanged by this column move. Once the rows are back in order, we can use the full column vectors to recover the original column sequence.

\subsection{Code Construction}
Let $\mathcal A_{q,n}$ denote the arrays in $\Sigma_q^{n\times n}$ with $\sum_{j=1}^n X_{i,j}\equiv 0\pmod q$ for every row $i$ and $\sum_{i=1}^n X_{i,j}\equiv 0\pmod q$ for every column $j$.

\begin{lemma}\label{lem:Aqn}
Let $q\geq2$ and $n\geq 3$. For each $r,c\in[1:n]$, the map $\Delta_{r,c}:\mathcal A_{q,n}\to\Sigma_q^{(n-1)\times(n-1)}$ is a bijection. In particular, $|\mathcal A_{q,n}|=q^{(n-1)^2}$.
\end{lemma}
\begin{IEEEproof}
Let $\bm Y$ be an arbitrary $(n-1)\times(n-1)$ array. Place it in the upper-left corner of $\bm X'$. Append the last column and then the last row by setting 
\begin{align}
X'_{i,n}&=-\sum_{j=1}^{n-1}Y_{i,j} \bmod{q}\qquad (i<n),\label{eq:parity-col}\\
X'_{n,j}&=-\sum_{i=1}^{n-1}X'_{i,j} \bmod{q} \qquad (j\leq n).\label{eq:parity-row}
\end{align}
This determines all the missing entries. Every column and each of the first $n-1$ rows now sum to zero modulo $q$. The last row therefore also sums to zero modulo $q$.  Moving the last row to position $r$ and the last column to position $c$ now gives the unique array in $\mathcal A_{q,n}$ with deletion output $\bm Y$ at $(r,c)$. Then the conclusion follows.
\end{IEEEproof}

The number of zeros in a row is already invariant under a column permutation. To distinguish rows with the same number of zeros, we group the columns by their numbers of zeros and count the zeros of the given row in each group. A column carries its group with it when it moves, so each zero is counted in the same group before and after the move. A column with no zeros contributes nothing. Permuting the rows also preserves the groups, since the number of zeros in each column stays the same.

\begin{definition}\label{def:labels}
For $\bm X\in\Sigma_q^{n\times n}$, let $d_j(\bm X)$ be the number of zeros in its $j$th column. The label $\lambda_i(\bm X)\in\Sigma_{n+1}^n$ of row $i$ is given by
\[
\lambda_i(\bm X)_d=|\{j\in[1:n]:X_{i,j}=0,\ d_j(\bm X)=d\}|
\]
for $d\in[1:n]$. Let $\Rank(\lambda(\bm X))$ denote the sequence $(\Rank(\lambda_i(\bm X)))_{i=1}^{n}$.
\end{definition}

Consider the binary array
\[
\bm X=\begin{bmatrix}
0&1&1&1&1\\
1&0&1&1&1\\
1&0&0&1&0\\
1&0&1&1&1\\
1&1&1&0&1
\end{bmatrix}.
\]
Each of the first two rows has one zero. These zeros lie in columns
containing one and three zeros, respectively. Thus
$\lambda_1(\bm X)=(1,0,0,0,0)$ and $\lambda_2(\bm X)=(0,0,1,0,0)$.
If we move column $1$ to the end, the two zeros move to columns $5$
and $1$, respectively. Each zero moves with its column, so it still
contributes to the same coordinate of its row label.

\begin{lemma}\label{lem:invariant}
Let $\rho,\kappa$ be permutations of $[1:n]$ and let $X'_{i,j}=X_{\rho(i),\kappa(j)}$. Then $d_j(\bm X')=d_{\kappa(j)}(\bm X)$ and $\lambda_i(\bm X')=\lambda_{\rho(i)}(\bm X)$.
\end{lemma}
\begin{IEEEproof}
A row permutation preserves the number of zeros in each column. Thus $d_j(\bm X')=d_{\kappa(j)}(\bm X)$. For each $d$, the number of $j$ such that $X_{\rho(i),\kappa(j)}=0$ and $d_{\kappa(j)}(\bm X)=d$ equals $\lambda_{\rho(i)}(\bm X)_d$, since $\kappa$ is a permutation. Thus, $\lambda_i(\bm X')=\lambda_{\rho(i)}(\bm X)$. This completes the proof.
\end{IEEEproof}

Once each column is available in full, compute the labels after parity completion. By Lemma~\ref{lem:invariant}, $\bm X'$ has the original label sequence with just the label of row $i^*$ moved to the end. We may take the first $n-1$ labels as a single-deletion output and the last label as the known deleted symbol. If adjacent rows share a label, recovering the label sequence can leave several possible positions for the missing row. We therefore require adjacent row labels to be distinct to determine the deleted row position. Once the row order has been restored, recovering the original column sequence suffices to recover the array, even if adjacent columns are equal.

\begin{construction}\label{cons:dense}
Let $\mathcal B_{q,n}$ consist of the arrays $\bm X\in\mathcal A_{q,n}$ such that
\begin{equation}\label{eq:Bqn}
\lambda_i(\bm X)\ne\lambda_{i+1}(\bm X)
\quad (i\in[1:n-1]).
\end{equation}
For $a,b\in[0:n-1]$, define $\mathcal C^1_{q,n}(a,b)$ to be the set of $\bm X\in\mathcal B_{q,n}$ satisfying
\begin{equation}\label{eq:code1}
\VTSyn(\Rank(\lambda(\bm X)))=a,
\qquad
\VTSyn(\Rank^c(\bm X))=b.
\end{equation}
\end{construction}

\begin{remark}
The sum of the coordinates of $\lambda_i(\bm X)$ equals the number of zeros in row $i$. For $q=2$, this count determines the row composition. Thus two rows with different compositions have different labels, and two rows with the same composition can have different labels if their zeros are distributed differently among the column groups. Requiring adjacent rows to have distinct labels is therefore weaker than requiring distinct compositions. This gives a larger set of admissible arrays and allows us to improve the redundancy achieved in~\cite{sun2026}.
\end{remark}

\subsection{Decoding}
\begin{figure}[!t]
\centering
\begin{minipage}[c]{0.46\linewidth}
\centering
\resizebox{\linewidth}{!}{$
\bm X=\begin{bmatrix}
X_{1,1}&\cdots&X_{1,j^*-1}&\textcolor{red}{X_{1,j^*}}&X_{1,j^*+1}&\cdots&X_{1,n}\\
\vdots&\ddots&\vdots&\textcolor{red}{\vdots}&\vdots&\ddots&\vdots\\
X_{i^*-1,1}&\cdots&X_{i^*-1,j^*-1}&\textcolor{red}{X_{i^*-1,j^*}}&X_{i^*-1,j^*+1}&\cdots&X_{i^*-1,n}\\
\textcolor{red}{X_{i^*,1}}&\textcolor{red}{\cdots}&\textcolor{red}{X_{i^*,j^*-1}}&\textcolor{red}{X_{i^*,j^*}}&\textcolor{red}{X_{i^*,j^*+1}}&\textcolor{red}{\cdots}&\textcolor{red}{X_{i^*,n}}\\
X_{i^*+1,1}&\cdots&X_{i^*+1,j^*-1}&\textcolor{red}{X_{i^*+1,j^*}}&X_{i^*+1,j^*+1}&\cdots&X_{i^*+1,n}\\
\vdots&\ddots&\vdots&\textcolor{red}{\vdots}&\vdots&\ddots&\vdots\\
X_{n,1}&\cdots&X_{n,j^*-1}&\textcolor{red}{X_{n,j^*}}&X_{n,j^*+1}&\cdots&X_{n,n}
\end{bmatrix}
$}
\end{minipage}\hfill
\begin{minipage}[c]{0.46\linewidth}
\centering
\resizebox{\linewidth}{!}{$
\bm Y=\begin{bmatrix}
X_{1,1}&\cdots&X_{1,j^*-1}&X_{1,j^*+1}&\cdots&X_{1,n}\\
\vdots&\ddots&\vdots&\vdots&\ddots&\vdots\\
X_{i^*-1,1}&\cdots&X_{i^*-1,j^*-1}&X_{i^*-1,j^*+1}&\cdots&X_{i^*-1,n}\\
X_{i^*+1,1}&\cdots&X_{i^*+1,j^*-1}&X_{i^*+1,j^*+1}&\cdots&X_{i^*+1,n}\\
\vdots&\ddots&\vdots&\vdots&\ddots&\vdots\\
X_{n,1}&\cdots&X_{n,j^*-1}&X_{n,j^*+1}&\cdots&X_{n,n}
\end{bmatrix}
$}
\end{minipage}
\par\medskip
\begin{minipage}[c]{0.46\linewidth}
\centering
\resizebox{\linewidth}{!}{$
\bm X'=\begin{bmatrix}
X_{1,1}&\cdots&X_{1,j^*-1}&X_{1,j^*+1}&\cdots&X_{1,n}&\textcolor{red}{X_{1,j^*}}\\
\vdots&\ddots&\vdots&\vdots&\ddots&\vdots&\textcolor{red}{\vdots}\\
X_{i^*-1,1}&\cdots&X_{i^*-1,j^*-1}&X_{i^*-1,j^*+1}&\cdots&X_{i^*-1,n}&\textcolor{red}{X_{i^*-1,j^*}}\\
X_{i^*+1,1}&\cdots&X_{i^*+1,j^*-1}&X_{i^*+1,j^*+1}&\cdots&X_{i^*+1,n}&\textcolor{red}{X_{i^*+1,j^*}}\\
\vdots&\ddots&\vdots&\vdots&\ddots&\vdots&\textcolor{red}{\vdots}\\
X_{n,1}&\cdots&X_{n,j^*-1}&X_{n,j^*+1}&\cdots&X_{n,n}&\textcolor{red}{X_{n,j^*}}\\
\textcolor{red}{X_{i^*,1}}&\textcolor{red}{\cdots}&\textcolor{red}{X_{i^*,j^*-1}}&\textcolor{red}{X_{i^*,j^*+1}}&\textcolor{red}{\cdots}&\textcolor{red}{X_{i^*,n}}&\textcolor{red}{X_{i^*,j^*}}
\end{bmatrix}
$}
\end{minipage}\hfill
\begin{minipage}[c]{0.46\linewidth}
\centering
\resizebox{\linewidth}{!}{$
\bm X''=\begin{bmatrix}
X_{1,1}&\cdots&X_{1,j^*-1}&X_{1,j^*+1}&\cdots&X_{1,n}&\textcolor{red}{X_{1,j^*}}\\
\vdots&\ddots&\vdots&\vdots&\ddots&\vdots&\textcolor{red}{\vdots}\\
X_{i^*-1,1}&\cdots&X_{i^*-1,j^*-1}&X_{i^*-1,j^*+1}&\cdots&X_{i^*-1,n}&\textcolor{red}{X_{i^*-1,j^*}}\\
\textcolor{red}{X_{i^*,1}}&\textcolor{red}{\cdots}&\textcolor{red}{X_{i^*,j^*-1}}&\textcolor{red}{X_{i^*,j^*+1}}&\textcolor{red}{\cdots}&\textcolor{red}{X_{i^*,n}}&\textcolor{red}{X_{i^*,j^*}}\\
X_{i^*+1,1}&\cdots&X_{i^*+1,j^*-1}&X_{i^*+1,j^*+1}&\cdots&X_{i^*+1,n}&\textcolor{red}{X_{i^*+1,j^*}}\\
\vdots&\ddots&\vdots&\vdots&\ddots&\vdots&\textcolor{red}{\vdots}\\
X_{n,1}&\cdots&X_{n,j^*-1}&X_{n,j^*+1}&\cdots&X_{n,n}&\textcolor{red}{X_{n,j^*}}
\end{bmatrix}
$}
\end{minipage}
\caption{Arrays used in the first decoder. The deleted row and column
are marked in red. Completing $\bm Y$ gives $\bm X'$, and restoring the
row order gives $\bm X''$.}
\label{fig:four_matrices}
\end{figure}
\begin{theorem}\label{thm:dec_c1}
Let $q\geq 2$, $n\geq 3$, and $a,b\in[0:n-1]$. Every nonempty code $\mathcal C^1_{q,n}(a,b)$ in Construction~\ref{cons:dense} corrects a $(1,1)$-criss-cross deletion. The codeword can be recovered in $O_q(n^2)$ time.
\end{theorem}

\begin{IEEEproof}
Suppose $\bm Y=\Delta_{i^*,j^*}(\bm X)$ for $\bm X\in\mathcal C^1_{q,n}(a,b)$. By Lemma~\ref{lem:Aqn}, we obtain $\bm X'$ by completing the last column and last row of $\bm Y$ according to~\eqref{eq:parity-col} and~\eqref{eq:parity-row}. This is the array obtained by moving row $i^*$ and column $j^*$ of $\bm X$ to the last positions. The four arrays in this procedure are shown in Fig.~\ref{fig:four_matrices}.

We first restore the row order. By Lemma~\ref{lem:invariant}, the labels of the first $n-1$ rows of $\bm X'$ are the labels of $\bm X$ with the $i^*$th entry deleted. The missing label is $\lambda_n(\bm X')=\lambda_{i^*}(\bm X)$. By~\eqref{eq:Bqn}, the adjacent row labels of $\bm X$ are distinct. Applying Corollary~\ref{cor:known-symbol} with the row syndrome $a$ in~\eqref{eq:code1} therefore gives $i^*$. Move the last row of $\bm X'$ to position $i^*$ and denote the resulting array by $\bm X''$.

After restoring the rows, the first $n-1$ columns of $\bm X''$ give the column sequence of $\bm X$ with one column deleted.
The last column of $\bm X''$ gives the value of the deleted column.
Applying Corollary~\ref{cor:known-symbol} with syndrome $b$ recovers the original sequence of columns and hence $\bm X$.

The parity completion takes $O_q(n^2)$ time. To compute the labels, we first count the zeros in each column of $\bm X'$. For every zero entry $X'_{i,j}$, we increment coordinate $d_j(\bm X')$ of $\lambda_i(\bm X')$. This takes $O_q(n^2)$ time, including initialization. By Corollary~\ref{cor:known-symbol}, each VT decoding step uses $O(n)$ lexicographic comparisons of length-$n$ vectors. Comparing two such vectors takes $O(n)$ time, so each VT decoding step takes $O(n^2)$ time. Restoring the row order and writing the recovered array also take $O(n^2)$ time. This gives the stated bound.
\end{IEEEproof}

\subsection{Cardinality and Redundancy}

The two syndromes partition $\mathcal B_{q,n}$ into at most $n^2$
nonempty classes. To bound the size of the largest class, we estimate
the probability that two adjacent rows of a uniformly chosen array in
$\mathcal A_{q,n}$ have the same label. A column in which exactly one
of the two rows has a zero contributes $1$ or $-1$ to the coordinate
given by its zero count. For the labels to agree, these contributions
must cancel separately in each coordinate.

\begin{lemma}\label{lem:Blabel}
Let $q\geq 2$ and $n\geq 3+(8q\log_2 15)^2$, and let $\bm X$ be
uniformly chosen from $\mathcal A_{q,n}$. For any two distinct rows
$i_1,i_2\in[1:n]$, we have
\begin{equation}\label{eq:label-collision}
\Pr\big(\lambda_{i_1}(\bm X)=\lambda_{i_2}(\bm X)\big)
\leq 15\cdot 2^{-\sqrt{n-2}/(8q)}.
\end{equation}
\end{lemma}
\begin{IEEEproof}
Choose a row $r\in[1:n]\setminus\{i_1,i_2\}$ and a column $c\in[1:n]$.
By Lemma~\ref{lem:Aqn}, the entries outside row $r$ and column $c$ are
independent and uniform. For $j\ne c$, let $I_j$ be the indicator that
exactly one of $X_{i_1,j}$ and $X_{i_2,j}$ is zero. Then
$\Pr(I_j=1)=2(q-1)/q^2$. Set $I=\sum_{j\ne c} I_j$, then $I\thicksim \mathrm{B}(n-1,2(q-1)/q^2).$
By Lemma~\ref{lem:markov} and independence,
\begin{align*}
    \Pr(I<\frac{(q-1)(n-1)}{q^2})
    &\leq \mathrm{exp}(\frac{(q-1)(n-1)}{q^2}) \mathbb{E}[e^{-I}]
    = \mathrm{exp}(\frac{(q-1)(n-1)}{q^2})(1-\frac{2(1-e^{-1})(q-1)}{q^2})^{n-1}\\
    &\leq \mathrm{exp}(-\frac{(q-1)(n-1)}{4q^2}) 
    \leq 2^{-\sqrt{n-2}/(8q)},
\end{align*}
where the second inequality holds because $1-x\leq e^{-x}$ and $1-2/e>1/4$, and the last inequality holds because $n-1\geq \sqrt{n-2}$ and $2(q-1)\geq q\ln2$.

For each $d\in[1:n]$, let $B_d$ count the columns $j\ne c$ for which
$I_j=1$ and $d_j(\bm X)=d$. Note that $H_j=\sum_{i\in [1:n]\setminus\{i_1,i_2,r\}}\bm{1}_{X_{i,j}=0}$ has distribution
$\mathrm B(n-3,1/q)$ and is independent of $I_j$. On $\{I_j=1\}$, 
we have $d_j(\bm X)=H_j+1+\ind_{\{X_{r,j}=0\}}$, so Lemma~\ref{lem:bin} and the union bound give
\begin{align*}
\Pr(I_j=1,\ d_j(\bm X)=d)
&=\mathrm{Pr}(I_j=1)\mathrm{Pr}(d_j(\bm{X})=d\vert I_j=1)
\leq \mathrm{Pr}(I_j=1) \left(\mathrm{Pr}(H_j=d-1)+\mathrm{Pr}(H_j=d-2)\right)\\
&\leq \frac{2(q-1)}{q^2}\frac{2q}{\sqrt{n-2}}
=\frac{4(q-1)}{q\sqrt{n-2}}.
\end{align*}
For fixed $d$, the events counted by $B_d$ are independent over
$j\ne c$, since $X_{r,j}$ is determined by the other entries in
column $j$. Applying
Lemma~\ref{lem:markov} to $B_d$ gives
\begin{align*}
    \Pr(B_d> \frac{8(q-1)(n-1)}{q\sqrt{n-2}})
    &\leq \mathrm{exp}(-\frac{8(q-1)(n-1)}{q\sqrt{n-2}}) \mathbb{E}[e^{B_d}]  
    \leq \mathrm{exp}(-\frac{8(q-1)(n-1)}{q\sqrt{n-2}}) \left(1+\frac{4(e-1)(q-1)}{q\sqrt{n-2}}\right)^{n-1}\\ 
    &\leq \mathrm{exp}(-\frac{4(3-e)(q-1)(n-1)}{q\sqrt{n-2}})  
    \leq \frac{8}{n}\cdot 2^{-\sqrt{n-2}/(8q)},
\end{align*}
where the third inequality holds because $1+x\leq e^x$, and the last inequality holds because $4(3-e)(q-1)/q-\ln2/(8q)\geq 2(3-e)-\ln2/16> 1/2$, $(n-1)/\sqrt{n-2}\geq \sqrt{n-2}$ and $e^{\sqrt{n-2}/2}\geq 1+(\sqrt{n-2}/2)^2/2=1+(n-2)/8\geq n/8$. 

Let $E$ be the event that $I\geq (q-1)(n-1)/q^2$ and $B_d\leq 8(q-1)(n-1)/(q\sqrt{n-2})$ for every $d\in[1:n]$. By the union bound, $$\Pr(E^{\mathrm c})\leq \Pr(I<\frac{(q-1)(n-1)}{q^2})+\sum_{d=1}^n\Pr(B_d> \frac{8(q-1)(n-1)}{q\sqrt{n-2}})\leq 9\cdot 2^{-\sqrt{n-2}/(8q)}.$$
On $E$, the set $D=\{d:B_d>0\}$ satisfies $|D|\geq I/(\max_{d\in [1:n]} B_d)\geq \sqrt{n-2}/(8q)$.

Condition on the free entries outside rows $i_1$ and $i_2$.
For each $j\ne c$, also fix the ordered pair $(X_{i_1,j},X_{i_2,j})$
if $I_j=0$ and the unordered pair $\{0,a_j\}$ if $I_j=1$.
When $I_j=1$, the two orders of the pair are equally likely and
independent across columns. Exchanging the entries of the pair
preserves $X_{r,j}$ and the zero count of column $j$, so $E$ and $D$
are determined by this conditioning.

For each $d\in D$, the contribution of columns $j\ne c$ to coordinate
$d$ of $\lambda_{i_1}(\bm X)-\lambda_{i_2}(\bm X)$ is a sum of
$B_d$ independent signs. This sum is zero with probability at most
$1/2$, since fixing all but one sign leaves at most one choice for
the last sign. The sums for distinct coordinates are independent.

All entries of column $c$ outside rows $i_1$ and $i_2$ are fixed
under the conditioning. In particular, $X_{r,c}$ is determined by
the other entries of row $r$, which are fixed. Thus $d_c(\bm X)$
has at most three possible values. For any such value $h$, the
event that the labels are equal and $d_c(\bm X)=h$ requires the
sign sums to vanish in all coordinates of $D\setminus\{h\}$.
The probability of these sums vanishing is at most $2^{1-|D|}$
under the conditioning above. On $E$, summing over the possible
values of $h$ and using $|D|\geq \sqrt{n-2}/(8q)$ gives
\[
\Pr\big(\lambda_{i_1}(\bm X)=\lambda_{i_2}(\bm X),\,E\big)
\leq 3\cdot 2^{1-|D|}
\leq 6\cdot 2^{-\sqrt{n-2}/(8q)}.
\]
Together with the bound on $\Pr(E^{\mathrm c})$, this proves
\eqref{eq:label-collision}.
\end{IEEEproof}

\begin{theorem}\label{thm:Bsize}
Let $q\geq 2$ and $n\geq 3+[16q(9+2\log_2 q)]^2$. Set
$\delta_{q,n}=15(n-1)2^{-\sqrt{n-2}/(8q)}$. Then
$\delta_{q,n}<1$ and
\[
|\mathcal B_{q,n}|\geq q^{(n-1)^2}(1-\delta_{q,n}).
\]
\end{theorem}
\begin{IEEEproof}
Let $\bm X$ be uniformly chosen from $\mathcal A_{q,n}$.
By~\eqref{eq:Bqn}, $\bm X\notin\mathcal B_{q,n}$ exactly when
two adjacent rows have the same label. By the union bound,
Lemma~\ref{lem:Blabel} then gives
\[
\Pr(\bm X\notin\mathcal B_{q,n})
\leq \sum_{i=1}^{n-1}
\Pr\big(\lambda_i(\bm X)=\lambda_{i+1}(\bm X)\big)
\leq \delta_{q,n}.
\]
Since $|\mathcal A_{q,n}|=q^{(n-1)^2}$ by Lemma~\ref{lem:Aqn},
we obtain $|\mathcal B_{q,n}|\geq q^{(n-1)^2}(1-\delta_{q,n})$.

Write $x=\sqrt{n-2}/(8q)$. Then
$\delta_{q,n}=15(64q^2x^2+1)2^{-x}$, which is decreasing for
$x\geq 22$ since its logarithmic derivative is less than
$2/x-\ln 2$. Our assumption gives $x\geq 18+4\log_2 q\geq 22$.
Evaluating at $18+4\log_2 q$ yields
\[
\delta_{q,n}
\leq \frac{15}{16}\left(
\frac{(9+2\log_2 q)^2}{64q^2}+\frac{1}{16384q^4}
\right)<1,
\]
where the last inequality follows from
$9+2\log_2 q\leq 7+2q\leq 11q/2$ for $q\geq 2$.
\end{IEEEproof}

\begin{theorem}\label{thm:dense-redundancy}
Let $q,n$ satisfy the assumptions of Theorem~\ref{thm:Bsize}.
There are $a,b\in[0:n-1]$ such that
\begin{align*}
|\mathcal C^1_{q,n}(a,b)|
&\geq \frac{q^{(n-1)^2}}{n^2}(1-\delta_{q,n}),\\
r_{\mathcal C^1_{q,n}(a,b)}
&\leq 2n-1+2\log_q n+
\frac{\delta_{q,n}}{(1-\delta_{q,n})\ln q}.
\end{align*}
\end{theorem}
\begin{IEEEproof}
Each array in $\mathcal B_{q,n}$ belongs to exactly one class
$\mathcal C^1_{q,n}(a,b)$. Since there are $n^2$ possible pairs
$(a,b)$, some class has size at least $|\mathcal B_{q,n}|/n^2$.
The cardinality bound now follows from Theorem~\ref{thm:Bsize}.
Taking logarithms gives a redundancy of at most
$2n-1+2\log_q n-\log_q(1-\delta_{q,n})$. Applying
$-\ln(1-x)\leq x/(1-x)$ for $0<x<1$ gives the stated bound.
\end{IEEEproof}

\begin{corollary}\label{cor:dense-redundancy}
For each fixed $q\geq 2$ and all sufficiently large $n$, there exist
$(1,1)$-criss-cross deletion-correcting codes with redundancy
$2n-1+2\log_q n+o_q(1)$ and $O_q(n^2)$-time decoding.
\end{corollary}
\begin{IEEEproof}
For fixed $q$, we have $\delta_{q,n}\to 0$ as $n\to\infty$.
The upper bound on the redundancy follows from
Theorem~\ref{thm:dense-redundancy}, and the lower bound follows from
Corollary~\ref{cor:lowerbound}. The decoding time follows from
Theorem~\ref{thm:dec_c1}.
\end{IEEEproof}

\section{Construction of Codes With Efficient Encoders}\label{sec:efficient}

In the previous section, we obtained a $(1,1)$-criss-cross deletion-correcting code with asymptotically optimal redundancy by choosing a largest syndrome class. Here we give a construction that admits $O_q(n^2)$-time encoding and has redundancy
$2n+2\log_q n+O_q(1)$. For each fixed $q$, this redundancy is within an additive constant of the lower bound.

Let $q\geq 2$, $n\geq q^4$, and $k=\lceil\log_q n\rceil$. Set
\[
u=\lceil\log_q(k^2)\rceil,\qquad v=k+2,
\]
and assume $u\geq 3$. For each fixed $q$, these assumptions hold for all sufficiently large $n$. Since $k\geq 4$, we have $q^k\geq 2^k\geq k^2$, so $u\leq k$ and $2\leq v-u\leq k-1$. Also, $n\geq 16$ gives $v\leq\log_2 n+3\leq n/2$, and hence $n-v\geq v\geq 6$. The bounds $q^u\geq k^2\geq (v-u)^2$ and $q^v\geq q^2n$ will be used to bound the redundancy incurred by the distinctness constraints. If $\log_q n$ and $2\log_q\log_q n$ are integers, then
$u=2\log_q\log_q n$ and $v=\log_q n+2$.

\subsection{Code Construction}
Partition $\bm X\in\Sigma_q^{n\times n}$ after its first $v$ rows and columns. Further partition the upper-left block $\bm A$ after its first $u$ rows and columns. Thus $\bm S$ is $u\times u$, $\bm U$ is $u\times(v-u)$, $\bm V$ is $(v-u)\times u$, and $\bm F$ is $(v-u)\times(v-u)$. The sizes of $\bm B$, $\bm C$, and $\bm D$ are $v\times(n-v)$, $(n-v)\times v$, and $(n-v)\times(n-v)$, respectively. Fig.~\ref{fig:blocks} shows the partition and the boundary markers.
\begin{equation}\label{eq:blocks}
\bm X=\begin{bmatrix}\bm A&\bm B\\ \bm C&\bm D\end{bmatrix},
\qquad
\bm A=\begin{bmatrix}\bm S&\bm U\\ \bm V&\bm F\end{bmatrix}.
\end{equation}

\begin{figure}[!t]
\centering
\begin{tikzpicture}[x=.5cm,y=-.5cm,font=\footnotesize]
\fill[seedcolor] (0,0) rectangle (2,2);
\fill[innercolor] (2,0) rectangle (5,2);
\fill[innercolor] (0,2) rectangle (2,5);
\fill[outercolor] (5,0) rectangle (12.5,5);
\fill[outercolor] (0,5) rectangle (5,12.5);
\fill[paritycolor] (5,12.18) rectangle (12.5,12.5);
\fill[paritycolor] (12.18,5) rectangle (12.5,12.5);
\draw[thick] (0,0) rectangle (12.5,12.5);
\draw[thick] (5,0)--(5,12.5);
\draw[thick] (0,5)--(12.5,5);
\draw (2,0)--(2,5);
\draw (0,2)--(5,2);
\draw[densely dashed] (5,12.18)--(12.18,12.18)--(12.18,5);
\node[font=\Large] at (1,1) {$\bm S$};
\node[font=\Large] at (3.5,1) {$\bm U$};
\node[font=\Large] at (1,3.5) {$\bm V$};
\node[font=\Large] at (3.5,3.5) {$\bm F$};
\node[font=\Large] at (8.75,2.5) {$\bm B$};
\node[font=\Large] at (2.5,8.75) {$\bm C$};
\node[font=\Large] at (8.6,8.6) {$\bm D$};
\foreach \x in {.2,.6} {
  \node at (\x,1.7) {$0$};
  \node at (\x,2.3) {$\star$};
  \node at (\x,4.7) {$0$};
  \node at (\x,5.3) {$\star$};
}
\foreach \y in {.2,.6} {
  \node at (1.7,\y) {$0$};
  \node at (2.3,\y) {$\star$};
  \node at (4.7,\y) {$0$};
  \node at (5.3,\y) {$\star$};
}
\foreach \a/\b in {0/2,2/5,5/12.5} {
  \draw[thin] (\a,-.25)--(\a,-.45)--(\b,-.45)--(\b,-.25);
}
\node[anchor=south] at (1,-.5) {$u$};
\node[anchor=south] at (3.5,-.5) {$v-u$};
\node[anchor=south] at (8.75,-.5) {$n-v$};
\foreach \a/\b in {0/2,2/5,5/12.5} {
  \draw[thin] (-.15,\a)--(-.3,\a)--(-.3,\b)--(-.15,\b);
}
\node[anchor=east] at (-.4,1) {$u$};
\node[anchor=east] at (-.4,3.5) {$v-u$};
\node[anchor=east] at (-.4,8.75) {$n-v$};
\end{tikzpicture}
\caption{Block decomposition of $\bm X$. Each $\star$ denotes a nonzero entry. The four upper-left blocks form $\bm A$. The shaded last row and column of $\bm D$ complete the remaining row and column sums.}
\label{fig:blocks}
\end{figure}

The boundary markers distinguish the three deletion intervals $[1:u]$, $[u+1:v]$, and $[v+1:n]$. At each boundary $h\in\{u,v\}$, the column markers are zeros at column $h$ and nonzeros at column $h+1$ in the first two rows. The row markers are defined by transposition. The block constraints below impose these markers.

\begin{definition}\label{def:canonical}
Let $\mathcal S_{q,u}$ consist of the arrays $\bm S\in\Sigma_q^{u\times u}$ whose first two entries in the last row and last column are zero, whose row labels are pairwise distinct, and whose column labels are pairwise distinct. Thus the conditions on $\bm S$ are
\begin{equation}\label{eq:S}
\begin{gathered}
S_{1,u}=S_{2,u}=S_{u,1}=S_{u,2}=0,\\
\lambda_{i_1}(\bm S)\ne\lambda_{i_2}(\bm S),\qquad
\lambda_{i_1}(\bm S^{\mathrm T})\ne\lambda_{i_2}(\bm S^{\mathrm T})
\quad (i_1,i_2\in[1:u],\ i_1\ne i_2).
\end{gathered}
\end{equation}
Since the row labels of $\bm S$ are pairwise distinct, let $\pi_r(i)$ be the index of the row whose label is the $i$th smallest in lexicographic order, for $i\in[1:u]$. Likewise, let $\pi_c(i)$ be the index of the column whose label is the $i$th smallest in lexicographic order. Here the label of column $j$ is $\lambda_j(\bm S^{\mathrm T})$. 
\end{definition}

Require the columns of $\bm U$ to be pairwise distinct. In its first column, the top two entries are nonzero, and in its last column, the top two entries are zero. These are the markers at columns $u+1$ and $v$ of $\bm X$. Thus $\bm U$ satisfies
\begin{equation}\label{eq:U}
\begin{gathered}
U_{1,1},U_{2,1}\ne 0,\qquad U_{1,v-u}=U_{2,v-u}=0,\\
\bm U_{[1:u],j_1}\ne\bm U_{[1:u],j_2}
\quad (j_1,j_2\in[1:v-u],\ j_1\ne j_2).
\end{gathered}
\end{equation}

Require $\bm V^{\mathrm T}$ to satisfy~\eqref{eq:U}. The rows of $\bm V$ are therefore pairwise distinct. The first two entries of its first row are nonzero, and those of its last row are zero, giving the markers at rows $u+1$ and $v$ of $\bm X$. In terms of $\bm V$, the conditions are
\begin{equation}\label{eq:V}
\begin{gathered}
V_{1,1},V_{1,2}\ne 0,\qquad V_{v-u,1}=V_{v-u,2}=0,\\
\bm V_{i_1,[1:u]}\ne\bm V_{i_2,[1:u]}
\quad (i_1,i_2\in[1:v-u],\ i_1\ne i_2).
\end{gathered}
\end{equation}

The blocks $\bm S$ and $\bm U$ share the first $u$ rows of $\bm X$. We read each column of $\bm U$ in the row order $\pi_r$ determined by $\bm S$. Similarly, $\bm S$ and $\bm V$ share the first $u$ columns, so we read each row of $\bm V$ in the column order $\pi_c$, i.e., define
\[
\begin{aligned}
\widehat U_{i,j}&=U_{\pi_r(i),j} && (i\in[1:u],\ j\in[1:v-u]),\\
\widehat V_{i,j}&=V_{i,\pi_c(j)} && (i\in[1:v-u],\ j\in[1:u]).
\end{aligned}
\]

For $t\in[1:v-u]$, let $j_t$ be the index of the $t$th smallest column of $\widehat{\bm U}$ in lexicographic order, and let $i_t$ be the index of the $t$th smallest row of $\widehat{\bm V}$ in lexicographic order. The distinctness conditions make both indices unique. Column $j_t$ of $\bm U$ is at position $u+j_t$ in $\bm X$, and row $i_t$ of $\bm V$ is at position $u+i_t$ in $\bm X$. We therefore define the permutations $\pi_U$ and $\pi_V$ of $[u+1:v]$ by
\[
\pi_U(u+t)=u+j_t,\qquad
\pi_V(u+t)=u+i_t
\quad (t\in[1:v-u]).
\]
Concatenating the orders on $[1:u]$ and $[u+1:v]$ gives the canonical row and column orders on $[1:v]$,
\[
\sigma_r=(\pi_r,\pi_V),\qquad \sigma_c=(\pi_c,\pi_U).
\]
Thus $\sigma_r^{-1}(i)$ gives the identifier of the row at position $i$, and $\sigma_c^{-1}(j)$ gives the identifier of the column at position $j$. The orders are determined by $\bm S$, $\bm U$, and $\bm V$.

We use the same rule to read $\bm B$ and $\bm C$, with $\bm A$ supplying the orders. The blocks $\bm A$ and $\bm B$ share the first $v$ rows of $\bm X$, so we read each column of $\bm B$ in order $\sigma_r$, just as we used $\pi_r$ to read the columns of $\bm U$. Similarly, $\bm A$ and $\bm C$ share the first $v$ columns, so we read each row of $\bm C$ in order $\sigma_c$, just as we used $\pi_c$ to read the rows of $\bm V$. The resulting representations are
\[
\begin{aligned}
\widehat B_{i,j}&=B_{\sigma_r(i),j} && (i\in[1:v],\ j\in[1:n-v]),\\
\widehat C_{i,j}&=C_{i,\sigma_c(j)} && (i\in[1:n-v],\ j\in[1:v]).
\end{aligned}
\]
To form the sequences for the VT constraints, we list the column ranks of $\widehat{\bm B}$ in physical column order and the row ranks of $\widehat{\bm C}$ in physical row order.

The first column of $\bm B$ has nonzero entries in its first two positions, supplying the markers at column $v+1$ of $\bm X$. The sum of the column vectors of $\bm B$ cancels that of $\bm A$ modulo $q$, so the first $v$ rows of $\bm X$ have zero sums modulo $q$. Adjacent columns of $\bm B$ are distinct, allowing the decoder to locate the deleted column within $\bm B$. The column rank sequence of $\widehat{\bm B}$ shares a VT syndrome with $\sigma_c^{-1}$. The conditions are
\begin{equation}\label{eq:B}
\begin{gathered}
B_{1,1},B_{2,1}\ne 0,\qquad
\bm B_{[1:v],j}\ne\bm B_{[1:v],j+1}
\quad (j\in[1:n-v-1]),\\
\sum_{j=1}^{n-v} \bm B_{[1:v],j}\equiv-\sum_{j=1}^v \bm A_{[1:v],j}\pmod q,\\
\VTSyn(\Rank^c(\widehat{\bm B}))=\VTSyn(\sigma_c^{-1}).
\end{gathered}
\end{equation}

The conditions on $\bm C$ are the row counterparts of those on $\bm B$. The first two entries of its first row are nonzero and supply the markers at row $v+1$ of $\bm X$. The sum of the row vectors of $\bm C$ cancels that of $\bm A$ modulo $q$, so the first $v$ column sums of $\bm X$ are zero modulo $q$. Adjacent rows are distinct, and the row rank sequence of $\widehat{\bm C}$ shares a VT syndrome with $\sigma_r^{-1}$. The conditions are
\begin{equation}\label{eq:C}
\begin{gathered}
C_{1,1},C_{1,2}\ne 0,\qquad
\bm C_{i,[1:v]}\ne\bm C_{i+1,[1:v]}
\quad (i\in[1:n-v-1]),\\
\sum_{i=1}^{n-v} \bm C_{i,[1:v]}\equiv-\sum_{i=1}^v \bm A_{i,[1:v]}\pmod q,\\
\VTSyn(\Rank^r(\widehat{\bm C}))=\VTSyn(\sigma_r^{-1}).
\end{gathered}
\end{equation}
In~\eqref{eq:B} and~\eqref{eq:C}, the VT syndromes are compared as nonnegative integers. The syndrome of a length-$v$ sequence lies in $[0:v-1]$, which is contained in $[0:n-v-1]$ since $n-v\geq v$.

The row sums of $\bm D$ cancel those of $\bm C$, and its column sums cancel those of $\bm B$. Together with the conditions on $\bm B$ and $\bm C$, this gives zero row and column sums modulo $q$ throughout $\bm X$. These sums determine the missing entries once the decoder has chosen positions for the deleted row and column. We require
\begin{equation}\label{eq:D}
\begin{aligned}
\sum_{j=1}^{n-v} D_{i,j}&\equiv-\sum_{j=1}^v C_{i,j}\pmod q && (i\in[1:n-v]),\\
\sum_{i=1}^{n-v} D_{i,j}&\equiv-\sum_{i=1}^v B_{i,j}\pmod q && (j\in[1:n-v]).
\end{aligned}
\end{equation}

\begin{remark}\label{rmk}
    Let $\bm X$ be decomposed as in~\eqref{eq:blocks}. If it satisfies~\eqref{eq:B}, \eqref{eq:C}, and~\eqref{eq:D}, then the sum of each row and each column is zero modulo $q$.
\end{remark}

\begin{construction}\label{cons:efficient}
Let $\mathcal C^2_{q,n}$ consist of all $q$-ary arrays with the partition in~\eqref{eq:blocks} and the stated block constraints.
\[
\mathcal C^2_{q,n}
=\left\{\bm X\in\Sigma_q^{n\times n}\ \middle|\
\begin{array}{@{}l@{}}
\bm S\in\mathcal S_{q,u},\quad \bm F\in\Sigma_q^{(v-u)\times (v-u)},\\
\bm U,\ \bm V,\
\bm B,\ \bm C,\ \bm D\text{ satisfy }\eqref{eq:U},\ \eqref{eq:V},\ \eqref{eq:B},\ \eqref{eq:C},\ \eqref{eq:D}\text{, respectively}
\end{array}
\right\}.
\]
\end{construction}

\subsection{Decoding}
\begin{lemma}\label{lem:invariant2}
Let $\bm A$, $\bm B$, and $\bm C$ be blocks as in~\eqref{eq:blocks}, and suppose the distinctness conditions defining $\sigma_r$ and $\sigma_c$ hold. Let $\rho$ and $\kappa$ be permutations of $[1:v]$ that preserve $[1:u]$ and $[u+1:v]$, and let $\omega$ and $\tau$ be permutations of $[1:n-v]$. Suppose
\[
A'_{i,j}=A_{\rho(i),\kappa(j)},\qquad
B'_{i,j}=B_{\rho(i),\omega(j)},\qquad
C'_{i,j}=C_{\tau(i),\kappa(j)}.
\]
Then $\sigma_r'=\rho^{-1}\sigma_r$ and $\sigma_c'=\kappa^{-1}\sigma_c$. Moreover, $\widehat B'_{i,j}=\widehat B_{i,\omega(j)}$ and $\widehat C'_{i,j}=\widehat C_{\tau(i),j}$.
\end{lemma}

\begin{IEEEproof}
By Lemma~\ref{lem:invariant}, row $i$ of $\bm S'$ has the label of row $\rho(i)$ of $\bm S$. Thus the $t$th smallest row label moves from position $\pi_r(t)$ to position $\rho^{-1}(\pi_r(t))$. The same argument applies to the column labels. Since the labels are distinct, we have $\pi_r'=\rho^{-1}\pi_r$ and $\pi_c'=\kappa^{-1}\pi_c$ on $[1:u]$.

By the definitions of $\widehat{\bm U}'$ and $\widehat{\bm V}'$, and using $\rho(\pi_r'(i))=\pi_r(i)$ and $\kappa(\pi_c'(j))=\pi_c(j)$, we have
\[
\begin{aligned}
\widehat U'_{i,j}
&=U'_{\pi_r'(i),j}
 =A'_{\pi_r'(i),u+j}
 =A_{\rho(\pi_r'(i)),\,\kappa(u+j)}
 =U_{\pi_r(i),\,\kappa(u+j)-u}
 =\widehat U_{i,\,\kappa(u+j)-u},\\[2pt]
\widehat V'_{i,j}
&=V'_{i,\pi_c'(j)}
 =A'_{u+i,\pi_c'(j)}
 =A_{\rho(u+i),\,\kappa(\pi_c'(j))}
 =V_{\rho(u+i)-u,\,\pi_c(j)}
 =\widehat V_{\rho(u+i)-u,\,j}.
\end{aligned}
\]
The first identity shows that each column retains its lexicographic rank. In particular, the $t$th smallest column is the same vector in $\widehat{\bm U}$ and $\widehat{\bm U}'$. Its position in $\bm A$ is $\pi_U(u+t)$, so its position in $\bm A'$ is $\kappa^{-1}(\pi_U(u+t))$. This gives $\pi_U'=\kappa^{-1}\pi_U$. Applying the same argument to the rows of $\widehat{\bm V}$ gives $\pi_V'=\rho^{-1}\pi_V$. Combining the orders on $[1:u]$ and $[u+1:v]$, we obtain $\sigma_r'=\rho^{-1}\sigma_r$ and $\sigma_c'=\kappa^{-1}\sigma_c$.

Finally, by the definitions of $\widehat{\bm B}'$ and $\widehat{\bm C}'$, and using $\rho(\sigma_r'(i))=\sigma_r(i)$ and $\kappa(\sigma_c'(j))=\sigma_c(j)$, we have
\[
\begin{aligned}
\widehat B'_{i,j}
&=B'_{\sigma_r'(i),j}
 =B_{\rho(\sigma_r'(i)),\,\omega(j)}
 =B_{\sigma_r(i),\,\omega(j)}
 =\widehat B_{i,\omega(j)},\\[2pt]
\widehat C'_{i,j}
&=C'_{i,\sigma_c'(j)}
 =C_{\tau(i),\,\kappa(\sigma_c'(j))}
 =C_{\tau(i),\,\sigma_c(j)}
 =\widehat C_{\tau(i),j}.
\end{aligned}
\]
This completes the proof.
\end{IEEEproof}

\begin{theorem}\label{thm:dec_c2}
Every nonempty code $\mathcal C^2_{q,n}$ in Construction~\ref{cons:efficient} corrects a $(1,1)$-criss-cross deletion. The codeword and both deletion positions can be recovered in $O_q(n^2)$ time.
\end{theorem}
\begin{IEEEproof}
Let $\bm Y=\Delta_{i^*,j^*}(\bm X)$ for $\bm X\in\mathcal C^2_{q,n}$. We decode $\bm Y$ in three steps.
\begin{enumerate}
    \item Use the boundary markers to identify the intervals containing $i^*$ and $j^*$.
    \item Insert a blank row and column at the right endpoints of their respective intervals, and use the zero row and column sums modulo $q$ to fill in the missing entries.
    \item Use the VT syndromes to determine $i^*$ and $j^*$, and move the recovered row and column back to these positions.
\end{enumerate}

In Step~1), the first row of $\bm Y$ comes from row $1$ or row $2$ of $\bm X$. For $h\in\{u,v\}$, the entry $Y_{1,h}$ comes from column $h$ of $\bm X$ if $j^*>h$, and from column $h+1$ otherwise. In both of the first two rows of $\bm X$, the marker at column $h$ is zero and the marker at column $h+1$ is nonzero. Applying the same argument to the first column of $\bm Y$, we obtain
\[
Y_{1,h}=0\ \Longleftrightarrow\ j^*>h,\qquad
Y_{h,1}=0\ \Longleftrightarrow\ i^*>h.
\]
Thus the right endpoint $j'$ of the interval containing $j^*$ is given by
\[
j'=
\begin{cases}
u, & Y_{1,u}\ne 0,\\
v, & Y_{1,u}=0 \text{ and } Y_{1,v}\ne 0,\\
n, & Y_{1,u}=Y_{1,v}=0.
\end{cases}
\]
The tests on $Y_{u,1}$ and $Y_{v,1}$ give the right endpoint $i'$ of the interval containing $i^*$ in the same way.

In Step~2), we insert a blank row at $i'$ and a blank column at $j'$ in $\bm Y$. Let $\bm X'$ denote the resulting array after completion. For $i\ne i'$ and $j\ne j'$, the entry $X'_{i,j}$ is $Y_{i-\ind_{\{i>i'\}},\,j-\ind_{\{j>j'\}}}$. We first use the row sums to fill column $j'$ outside row $i'$, and then use the column sums to fill row $i'$, including the entry at $(i',j')$. Specifically, we set
\[
\begin{aligned}
X'_{i,j'}&\equiv-\sum_{k\ne j'}X'_{i,k}\pmod q
&& (i\ne i'),\\
X'_{i',j}&\equiv-\sum_{\ell\ne i'}X'_{\ell,j}\pmod q
&& (j\in[1:n]).
\end{aligned}
\]
These equations determine all missing entries. Moving row $i^*$ to $i'$ and column $j^*$ to $j'$ in $\bm X$, while preserving the order of the other rows and columns, gives an array with the same surviving entries and zero row and column sums. Hence $\bm X'$ is the array obtained by these two moves. Since each move is within its interval, the induced permutations satisfy the hypotheses of Lemma~\ref{lem:invariant2}.

In Step~3), we first recover $j^*$. Let $\kappa$ and $\omega$ be the column permutations in Lemma~\ref{lem:invariant2}, which map positions in $\bm X'$ to their original positions in the corresponding blocks. By the lemma, $(\sigma_c')^{-1}=\sigma_c^{-1}\circ\kappa$ and $\widehat B'_{i,j}=\widehat B_{i,\omega(j)}$. Thus
\[
\begin{aligned}
(\sigma_c')^{-1}(j)
&=\sigma_c^{-1}(\kappa(j))
&& (j\in[1:v]),\\
\Rank\bigl(\widehat{\bm B}'_{[1:v],j}\bigr)
&=\Rank\bigl(\widehat{\bm B}_{[1:v],\omega(j)}\bigr)
&& (j\in[1:n-v]).
\end{aligned}
\]
Therefore, each column retains its identifier or rank after the two moves. The symbol at the known position $j'$ represents the deleted column. Removing this symbol leaves the other symbols in their original order. The sequence for the other group of columns supplies the VT syndrome by~\eqref{eq:B}.

\begin{itemize}
    \item If $j'\le v$, the move is within the first $v$ columns. Thus $\omega$ is the identity and $\widehat{\bm B}'=\widehat{\bm B}$. The identifier at position $j^*$ of $\sigma_c^{-1}$ moves to position $j'$ of $(\sigma_c')^{-1}$. Hence
    \[
    \begin{aligned}
    (\sigma_c')^{-1}(j')&=\sigma_c^{-1}(j^*),\\
    \Delta_{j'}\bigl((\sigma_c')^{-1}\bigr)
        &=\Delta_{j^*}(\sigma_c^{-1}),\\
    \VTSyn\bigl(\Rank^c(\widehat{\bm B}')\bigr)
        &=\VTSyn(\sigma_c^{-1}).
    \end{aligned}
    \]
    The left-hand sides give the known symbol, the deletion subsequence, and the VT syndrome. Since the identifiers in $\sigma_c^{-1}$ are distinct, Corollary~\ref{cor:known-symbol} gives the original position $j^*$.

    \item If $j'=n$, the move is within the last $n-v$ columns. Thus $\kappa$ is the identity and $(\sigma_c')^{-1}=\sigma_c^{-1}$. The rank at position $j^*-v$ of $\Rank^c(\widehat{\bm B})$ moves to position $n-v$ of $\Rank^c(\widehat{\bm B}')$. Hence
    \[
    \begin{aligned}
    \Rank\bigl(\widehat{\bm B}'_{[1:v],n-v}\bigr)
        &=\Rank\bigl(\widehat{\bm B}_{[1:v],j^*-v}\bigr),\\
    \Delta_{n-v}\bigl(\Rank^c(\widehat{\bm B}')\bigr)
        &=\Delta_{j^*-v}\bigl(\Rank^c(\widehat{\bm B})\bigr),\\
    \VTSyn\bigl((\sigma_c')^{-1}\bigr)
        &=\VTSyn\bigl(\Rank^c(\widehat{\bm B})\bigr).
    \end{aligned}
    \]
    By~\eqref{eq:B}, adjacent columns of $\bm B$ are distinct. Reading both columns in the same row order preserves this property, so adjacent symbols of $\Rank^c(\widehat{\bm B})$ are distinct. Applying Corollary~\ref{cor:known-symbol} with the three quantities on the left gives the position $j^*-v$ within $\bm B$. Adding $v$ gives $j^*$.
\end{itemize}
The row position is recovered in the same way from $(\sigma_r')^{-1}$ and $\Rank^r(\widehat{\bm C}')$, using~\eqref{eq:C}. If $i'\le v$, we remove the identifier at position $i'$ of $(\sigma_r')^{-1}$ and use the syndrome of $\Rank^r(\widehat{\bm C}')$ to recover $i^*$. If $i'=n$, we remove the last symbol of $\Rank^r(\widehat{\bm C}')$ and use the syndrome of $(\sigma_r')^{-1}$ to recover $i^*-v$. Moving row $i'$ to $i^*$ and column $j'$ to $j^*$ gives $\bm X$.

Step~1) takes $O(1)$ time, since we only read four entries of $\bm Y$. Step~2) takes $O_q(n^2)$ time to compute the row and column sums and fill in the missing entries.
In Step~3), computing the labels of $\bm S'$ takes $O_q(u^2)$ time. Since each label has $u$ coordinates in $[0:u]$, we sort the labels in $O(u^2)$ time using stable counting sort. In the same way, sorting the columns of $\widehat{\bm U}'$ and the rows of $\widehat{\bm V}'$ takes $O(u(v-u))$ time. Each column of $\widehat{\bm B}'$ and each row of $\widehat{\bm C}'$ has $v$ entries, so computing the two rank sequences takes $O(v(n-v))$ time. Since $q^v=O_q(n)$, each rank fits in one word. We then compute the VT syndromes and apply Corollary~\ref{cor:known-symbol} in $O(n)$ time. Moving the recovered row and column back to their original positions takes $O_q(n^2)$ time. Hence Step~3) takes $O_q(n^2)$ time.
Combining the three steps, we obtain a decoding time of $O_q(n^2)$.
\end{IEEEproof}

\begin{example}\label{ex:decoding}
Let $\bm Y$ be the received array in Fig.~\ref{fig:example}, where $q=2$, $u=4$, $v=6$, and $n=16$.

In Step~1), $Y_{4,1}=1$ gives $i^*\in[1:4]$, and $Y_{1,6}=0$ gives $j^*\in[7:16]$. Thus $i'=4$ and $j'=16$.

In Step~2), we insert a blank row at position $4$ and a blank column at position $16$. For example, the entries in row $3$ of $\bm Y$ sum to $9$, so $X'_{3,16}\equiv -9\equiv 1\pmod 2$. After filling column $16$ outside row $4$, its entries sum to $7$. Hence $X'_{4,16}\equiv -7\equiv 1\pmod 2$. The other missing entries are filled in the same way using the row and column sums.

In Step~3), Lemma~\ref{lem:invariant2} allows us to recover the two positions independently from $\bm X'$. We first recover $i^*$. The sequences used for row recovery are
\[
\begin{aligned}
(\sigma_r')^{-1}&=(2,3,4,1,6,5),\\
\Rank^r(\widehat{\bm C}')&=(44,0,12,49,36,38,40,63,14,29).
\end{aligned}
\]
Since $i'=4\le v$, we remove the identifier $1$ at position $4$ of $(\sigma_r')^{-1}$ and use the VT syndrome of $\Rank^r(\widehat{\bm C}')$, which is $(2+3+5+6+7+9)\bmod 10=2$. Inserting $1$ at position $3$ of $(2,3,4,6,5)$ gives $(2,3,1,4,6,5)$, whose VT syndrome is $(1+3+4)\bmod 6=2$. Thus Corollary~\ref{cor:known-symbol} gives $i^*=3$.

We next recover $j^*$. Since $j'=n$, we use the last symbol of $\Rank^c(\widehat{\bm B}')$. Reading the last column of $\bm B'$ in the order $\sigma_r'=(4,1,2,3,6,5)$ gives $(1,0,0,1,1,0)^{\mathrm T}$, whose rank is $2^5+2^2+2^1=38$. Removing this symbol leaves $(30,39,21,29,33,63,18,46,5)$. The VT syndrome of $(\sigma_c')^{-1}=(3,1,2,4,6,5)$ is $(2+3+4)\bmod 6=3$. Inserting $38$ at position $4$ gives
$(30,39,21,38,29,33,63,18,46,5)$,
whose VT syndrome is $(1+3+5+6+8)\bmod 10=3$. Hence the original column position is $j^*=v+4=10$.

Finally, we recover $\bm X$ by moving row $4$ to position $3$ and column $16$ to position $10$ in $\bm X'$.
\end{example}

\begin{figure}[t!]
\centering
\begin{minipage}[t]{.32\linewidth}\centering\vspace{0pt}
\resizebox{\linewidth}{!}{%
\begin{tikzpicture}[x=.30cm,y=-.30cm,font=\scriptsize]
\tikzset{every node/.style={inner sep=0pt}}
\path[use as bounding box] (-2.2,-2.2) rectangle (16.3,16.3);
\fill[seedcolor] (0,0) rectangle (1,1);
\node at (0.5,0.5) {0};
\fill[seedcolor] (1,0) rectangle (2,1);
\node at (1.5,0.5) {1};
\fill[seedcolor] (2,0) rectangle (3,1);
\node at (2.5,0.5) {1};
\fill[seedcolor] (3,0) rectangle (4,1);
\node at (3.5,0.5) {0};
\fill[innercolor] (4,0) rectangle (5,1);
\node at (4.5,0.5) {1};
\fill[innercolor] (5,0) rectangle (6,1);
\node at (5.5,0.5) {0};
\fill[outercolor] (6,0) rectangle (7,1);
\node at (6.5,0.5) {1};
\fill[outercolor] (7,0) rectangle (8,1);
\node at (7.5,0.5) {0};
\fill[outercolor] (8,0) rectangle (9,1);
\node at (8.5,0.5) {1};
\fill[outercolor] (9,0) rectangle (10,1);
\node at (9.5,0.5) {0};
\fill[outercolor] (10,0) rectangle (11,1);
\node at (10.5,0.5) {1};
\fill[outercolor] (11,0) rectangle (12,1);
\node at (11.5,0.5) {0};
\fill[outercolor] (12,0) rectangle (13,1);
\node at (12.5,0.5) {1};
\fill[outercolor] (13,0) rectangle (14,1);
\node at (13.5,0.5) {1};
\fill[outercolor] (14,0) rectangle (15,1);
\node at (14.5,0.5) {0};
\fill[outercolor] (15,0) rectangle (16,1);
\node at (15.5,0.5) {0};
\fill[seedcolor] (0,1) rectangle (1,2);
\node at (0.5,1.5) {0};
\fill[seedcolor] (1,1) rectangle (2,2);
\node at (1.5,1.5) {1};
\fill[seedcolor] (2,1) rectangle (3,2);
\node at (2.5,1.5) {0};
\fill[seedcolor] (3,1) rectangle (4,2);
\node at (3.5,1.5) {0};
\fill[innercolor] (4,1) rectangle (5,2);
\node at (4.5,1.5) {1};
\fill[innercolor] (5,1) rectangle (6,2);
\node at (5.5,1.5) {0};
\fill[outercolor] (6,1) rectangle (7,2);
\node at (6.5,1.5) {1};
\fill[outercolor] (7,1) rectangle (8,2);
\node at (7.5,1.5) {0};
\fill[outercolor] (8,1) rectangle (9,2);
\node at (8.5,1.5) {0};
\fill[outercolor] (9,1) rectangle (10,2);
\node at (9.5,1.5) {0};
\fill[outercolor] (10,1) rectangle (11,2);
\node at (10.5,1.5) {1};
\fill[outercolor] (11,1) rectangle (12,2);
\node at (11.5,1.5) {0};
\fill[outercolor] (12,1) rectangle (13,2);
\node at (12.5,1.5) {1};
\fill[outercolor] (13,1) rectangle (14,2);
\node at (13.5,1.5) {0};
\fill[outercolor] (14,1) rectangle (15,2);
\node at (14.5,1.5) {1};
\fill[outercolor] (15,1) rectangle (16,2);
\node at (15.5,1.5) {0};
\fill[seedcolor] (0,2) rectangle (1,3);
\node at (0.5,2.5) {1};
\fill[seedcolor] (1,2) rectangle (2,3);
\node at (1.5,2.5) {1};
\fill[seedcolor] (2,2) rectangle (3,3);
\node at (2.5,2.5) {1};
\fill[seedcolor] (3,2) rectangle (4,3);
\node at (3.5,2.5) {0};
\fill[innercolor] (4,2) rectangle (5,3);
\node at (4.5,2.5) {0};
\fill[innercolor] (5,2) rectangle (6,3);
\node at (5.5,2.5) {0};
\fill[outercolor] (6,2) rectangle (7,3);
\node at (6.5,2.5) {0};
\fill[outercolor] (7,2) rectangle (8,3);
\node at (7.5,2.5) {1};
\fill[outercolor] (8,2) rectangle (9,3);
\node at (8.5,2.5) {0};
\fill[outercolor] (9,2) rectangle (10,3);
\node at (9.5,2.5) {1};
\fill[outercolor] (10,2) rectangle (11,3);
\node at (10.5,2.5) {0};
\fill[outercolor] (11,2) rectangle (12,3);
\node at (11.5,2.5) {1};
\fill[outercolor] (12,2) rectangle (13,3);
\node at (12.5,2.5) {1};
\fill[outercolor] (13,2) rectangle (14,3);
\node at (13.5,2.5) {0};
\fill[outercolor] (14,2) rectangle (15,3);
\node at (14.5,2.5) {1};
\fill[outercolor] (15,2) rectangle (16,3);
\node at (15.5,2.5) {0};
\fill[seedcolor] (0,3) rectangle (1,4);
\node at (0.5,3.5) {0};
\fill[seedcolor] (1,3) rectangle (2,4);
\node at (1.5,3.5) {0};
\fill[seedcolor] (2,3) rectangle (3,4);
\node at (2.5,3.5) {0};
\fill[seedcolor] (3,3) rectangle (4,4);
\node at (3.5,3.5) {0};
\fill[innercolor] (4,3) rectangle (5,4);
\node at (4.5,3.5) {1};
\fill[innercolor] (5,3) rectangle (6,4);
\node at (5.5,3.5) {1};
\fill[outercolor] (6,3) rectangle (7,4);
\node at (6.5,3.5) {1};
\fill[outercolor] (7,3) rectangle (8,4);
\node at (7.5,3.5) {1};
\fill[outercolor] (8,3) rectangle (9,4);
\node at (8.5,3.5) {1};
\fill[outercolor] (9,3) rectangle (10,4);
\node at (9.5,3.5) {1};
\fill[outercolor] (10,3) rectangle (11,4);
\node at (10.5,3.5) {1};
\fill[outercolor] (11,3) rectangle (12,4);
\node at (11.5,3.5) {0};
\fill[outercolor] (12,3) rectangle (13,4);
\node at (12.5,3.5) {1};
\fill[outercolor] (13,3) rectangle (14,4);
\node at (13.5,3.5) {0};
\fill[outercolor] (14,3) rectangle (15,4);
\node at (14.5,3.5) {1};
\fill[outercolor] (15,3) rectangle (16,4);
\node at (15.5,3.5) {1};
\fill[innercolor] (0,4) rectangle (1,5);
\node at (0.5,4.5) {1};
\fill[innercolor] (1,4) rectangle (2,5);
\node at (1.5,4.5) {1};
\fill[innercolor] (2,4) rectangle (3,5);
\node at (2.5,4.5) {0};
\fill[innercolor] (3,4) rectangle (4,5);
\node at (3.5,4.5) {1};
\node at (4.5,4.5) {1};
\node at (5.5,4.5) {0};
\fill[outercolor] (6,4) rectangle (7,5);
\node at (6.5,4.5) {0};
\fill[outercolor] (7,4) rectangle (8,5);
\node at (7.5,4.5) {1};
\fill[outercolor] (8,4) rectangle (9,5);
\node at (8.5,4.5) {1};
\fill[outercolor] (9,4) rectangle (10,5);
\node at (9.5,4.5) {0};
\fill[outercolor] (10,4) rectangle (11,5);
\node at (10.5,4.5) {1};
\fill[outercolor] (11,4) rectangle (12,5);
\node at (11.5,4.5) {1};
\fill[outercolor] (12,4) rectangle (13,5);
\node at (12.5,4.5) {1};
\fill[outercolor] (13,4) rectangle (14,5);
\node at (13.5,4.5) {0};
\fill[outercolor] (14,4) rectangle (15,5);
\node at (14.5,4.5) {0};
\fill[outercolor] (15,4) rectangle (16,5);
\node at (15.5,4.5) {1};
\fill[innercolor] (0,5) rectangle (1,6);
\node at (0.5,5.5) {0};
\fill[innercolor] (1,5) rectangle (2,6);
\node at (1.5,5.5) {0};
\fill[innercolor] (2,5) rectangle (3,6);
\node at (2.5,5.5) {1};
\fill[innercolor] (3,5) rectangle (4,6);
\node at (3.5,5.5) {0};
\node at (4.5,5.5) {1};
\node at (5.5,5.5) {0};
\fill[outercolor] (6,5) rectangle (7,6);
\node at (6.5,5.5) {1};
\fill[outercolor] (7,5) rectangle (8,6);
\node at (7.5,5.5) {1};
\fill[outercolor] (8,5) rectangle (9,6);
\node at (8.5,5.5) {0};
\fill[outercolor] (9,5) rectangle (10,6);
\node at (9.5,5.5) {1};
\fill[outercolor] (10,5) rectangle (11,6);
\node at (10.5,5.5) {0};
\fill[outercolor] (11,5) rectangle (12,6);
\node at (11.5,5.5) {0};
\fill[outercolor] (12,5) rectangle (13,6);
\node at (12.5,5.5) {1};
\fill[outercolor] (13,5) rectangle (14,6);
\node at (13.5,5.5) {1};
\fill[outercolor] (14,5) rectangle (15,6);
\node at (14.5,5.5) {1};
\fill[outercolor] (15,5) rectangle (16,6);
\node at (15.5,5.5) {0};
\fill[outercolor] (0,6) rectangle (1,7);
\node at (0.5,6.5) {1};
\fill[outercolor] (1,6) rectangle (2,7);
\node at (1.5,6.5) {1};
\fill[outercolor] (2,6) rectangle (3,7);
\node at (2.5,6.5) {0};
\fill[outercolor] (3,6) rectangle (4,7);
\node at (3.5,6.5) {1};
\fill[outercolor] (4,6) rectangle (5,7);
\node at (4.5,6.5) {0};
\fill[outercolor] (5,6) rectangle (6,7);
\node at (5.5,6.5) {0};
\node at (6.5,6.5) {1};
\node at (7.5,6.5) {0};
\node at (8.5,6.5) {1};
\node at (9.5,6.5) {0};
\node at (10.5,6.5) {1};
\node at (11.5,6.5) {0};
\node at (12.5,6.5) {1};
\node at (13.5,6.5) {1};
\node at (14.5,6.5) {0};
\fill[paritycolor] (15,6) rectangle (16,7);
\node at (15.5,6.5) {0};
\fill[outercolor] (0,7) rectangle (1,8);
\node at (0.5,7.5) {0};
\fill[outercolor] (1,7) rectangle (2,8);
\node at (1.5,7.5) {0};
\fill[outercolor] (2,7) rectangle (3,8);
\node at (2.5,7.5) {0};
\fill[outercolor] (3,7) rectangle (4,8);
\node at (3.5,7.5) {0};
\fill[outercolor] (4,7) rectangle (5,8);
\node at (4.5,7.5) {0};
\fill[outercolor] (5,7) rectangle (6,8);
\node at (5.5,7.5) {0};
\node at (6.5,7.5) {0};
\node at (7.5,7.5) {0};
\node at (8.5,7.5) {1};
\node at (9.5,7.5) {0};
\node at (10.5,7.5) {1};
\node at (11.5,7.5) {1};
\node at (12.5,7.5) {1};
\node at (13.5,7.5) {1};
\node at (14.5,7.5) {0};
\fill[paritycolor] (15,7) rectangle (16,8);
\node at (15.5,7.5) {1};
\fill[outercolor] (0,8) rectangle (1,9);
\node at (0.5,8.5) {1};
\fill[outercolor] (1,8) rectangle (2,9);
\node at (1.5,8.5) {0};
\fill[outercolor] (2,8) rectangle (3,9);
\node at (2.5,8.5) {0};
\fill[outercolor] (3,8) rectangle (4,9);
\node at (3.5,8.5) {1};
\fill[outercolor] (4,8) rectangle (5,9);
\node at (4.5,8.5) {0};
\fill[outercolor] (5,8) rectangle (6,9);
\node at (5.5,8.5) {0};
\node at (6.5,8.5) {0};
\node at (7.5,8.5) {0};
\node at (8.5,8.5) {0};
\node at (9.5,8.5) {0};
\node at (10.5,8.5) {1};
\node at (11.5,8.5) {1};
\node at (12.5,8.5) {0};
\node at (13.5,8.5) {0};
\node at (14.5,8.5) {1};
\fill[paritycolor] (15,8) rectangle (16,9);
\node at (15.5,8.5) {1};
\fill[outercolor] (0,9) rectangle (1,10);
\node at (0.5,9.5) {0};
\fill[outercolor] (1,9) rectangle (2,10);
\node at (1.5,9.5) {1};
\fill[outercolor] (2,9) rectangle (3,10);
\node at (2.5,9.5) {1};
\fill[outercolor] (3,9) rectangle (4,10);
\node at (3.5,9.5) {0};
\fill[outercolor] (4,9) rectangle (5,10);
\node at (4.5,9.5) {1};
\fill[outercolor] (5,9) rectangle (6,10);
\node at (5.5,9.5) {0};
\node at (6.5,9.5) {1};
\node at (7.5,9.5) {1};
\node at (8.5,9.5) {0};
\node at (9.5,9.5) {1};
\node at (10.5,9.5) {0};
\node at (11.5,9.5) {0};
\node at (12.5,9.5) {1};
\node at (13.5,9.5) {0};
\node at (14.5,9.5) {1};
\fill[paritycolor] (15,9) rectangle (16,10);
\node at (15.5,9.5) {0};
\fill[outercolor] (0,10) rectangle (1,11);
\node at (0.5,10.5) {0};
\fill[outercolor] (1,10) rectangle (2,11);
\node at (1.5,10.5) {1};
\fill[outercolor] (2,10) rectangle (3,11);
\node at (2.5,10.5) {0};
\fill[outercolor] (3,10) rectangle (4,11);
\node at (3.5,10.5) {1};
\fill[outercolor] (4,10) rectangle (5,11);
\node at (4.5,10.5) {0};
\fill[outercolor] (5,10) rectangle (6,11);
\node at (5.5,10.5) {0};
\node at (6.5,10.5) {0};
\node at (7.5,10.5) {1};
\node at (8.5,10.5) {0};
\node at (9.5,10.5) {1};
\node at (10.5,10.5) {1};
\node at (11.5,10.5) {0};
\node at (12.5,10.5) {0};
\node at (13.5,10.5) {1};
\node at (14.5,10.5) {1};
\fill[paritycolor] (15,10) rectangle (16,11);
\node at (15.5,10.5) {1};
\fill[outercolor] (0,11) rectangle (1,12);
\node at (0.5,11.5) {0};
\fill[outercolor] (1,11) rectangle (2,12);
\node at (1.5,11.5) {1};
\fill[outercolor] (2,11) rectangle (3,12);
\node at (2.5,11.5) {0};
\fill[outercolor] (3,11) rectangle (4,12);
\node at (3.5,11.5) {1};
\fill[outercolor] (4,11) rectangle (5,12);
\node at (4.5,11.5) {0};
\fill[outercolor] (5,11) rectangle (6,12);
\node at (5.5,11.5) {1};
\node at (6.5,11.5) {1};
\node at (7.5,11.5) {0};
\node at (8.5,11.5) {1};
\node at (9.5,11.5) {0};
\node at (10.5,11.5) {0};
\node at (11.5,11.5) {1};
\node at (12.5,11.5) {0};
\node at (13.5,11.5) {1};
\node at (14.5,11.5) {0};
\fill[paritycolor] (15,11) rectangle (16,12);
\node at (15.5,11.5) {1};
\fill[outercolor] (0,12) rectangle (1,13);
\node at (0.5,12.5) {1};
\fill[outercolor] (1,12) rectangle (2,13);
\node at (1.5,12.5) {1};
\fill[outercolor] (2,12) rectangle (3,13);
\node at (2.5,12.5) {0};
\fill[outercolor] (3,12) rectangle (4,13);
\node at (3.5,12.5) {0};
\fill[outercolor] (4,12) rectangle (5,13);
\node at (4.5,12.5) {0};
\fill[outercolor] (5,12) rectangle (6,13);
\node at (5.5,12.5) {0};
\node at (6.5,12.5) {0};
\node at (7.5,12.5) {1};
\node at (8.5,12.5) {1};
\node at (9.5,12.5) {1};
\node at (10.5,12.5) {0};
\node at (11.5,12.5) {1};
\node at (12.5,12.5) {0};
\node at (13.5,12.5) {0};
\node at (14.5,12.5) {0};
\fill[paritycolor] (15,12) rectangle (16,13);
\node at (15.5,12.5) {0};
\fill[outercolor] (0,13) rectangle (1,14);
\node at (0.5,13.5) {1};
\fill[outercolor] (1,13) rectangle (2,14);
\node at (1.5,13.5) {1};
\fill[outercolor] (2,13) rectangle (3,14);
\node at (2.5,13.5) {1};
\fill[outercolor] (3,13) rectangle (4,14);
\node at (3.5,13.5) {1};
\fill[outercolor] (4,13) rectangle (5,14);
\node at (4.5,13.5) {1};
\fill[outercolor] (5,13) rectangle (6,14);
\node at (5.5,13.5) {1};
\node at (6.5,13.5) {1};
\node at (7.5,13.5) {1};
\node at (8.5,13.5) {0};
\node at (9.5,13.5) {0};
\node at (10.5,13.5) {1};
\node at (11.5,13.5) {0};
\node at (12.5,13.5) {1};
\node at (13.5,13.5) {0};
\node at (14.5,13.5) {1};
\fill[paritycolor] (15,13) rectangle (16,14);
\node at (15.5,13.5) {1};
\fill[outercolor] (0,14) rectangle (1,15);
\node at (0.5,14.5) {1};
\fill[outercolor] (1,14) rectangle (2,15);
\node at (1.5,14.5) {0};
\fill[outercolor] (2,14) rectangle (3,15);
\node at (2.5,14.5) {0};
\fill[outercolor] (3,14) rectangle (4,15);
\node at (3.5,14.5) {1};
\fill[outercolor] (4,14) rectangle (5,15);
\node at (4.5,14.5) {0};
\fill[outercolor] (5,14) rectangle (6,15);
\node at (5.5,14.5) {1};
\node at (6.5,14.5) {0};
\node at (7.5,14.5) {0};
\node at (8.5,14.5) {1};
\node at (9.5,14.5) {1};
\node at (10.5,14.5) {0};
\node at (11.5,14.5) {0};
\node at (12.5,14.5) {1};
\node at (13.5,14.5) {1};
\node at (14.5,14.5) {1};
\fill[paritycolor] (15,14) rectangle (16,15);
\node at (15.5,14.5) {0};
\fill[outercolor] (0,15) rectangle (1,16);
\node at (0.5,15.5) {1};
\fill[outercolor] (1,15) rectangle (2,16);
\node at (1.5,15.5) {0};
\fill[outercolor] (2,15) rectangle (3,16);
\node at (2.5,15.5) {1};
\fill[outercolor] (3,15) rectangle (4,16);
\node at (3.5,15.5) {1};
\fill[outercolor] (4,15) rectangle (5,16);
\node at (4.5,15.5) {1};
\fill[outercolor] (5,15) rectangle (6,16);
\node at (5.5,15.5) {0};
\fill[paritycolor] (6,15) rectangle (7,16);
\node at (6.5,15.5) {0};
\fill[paritycolor] (7,15) rectangle (8,16);
\node at (7.5,15.5) {0};
\fill[paritycolor] (8,15) rectangle (9,16);
\node at (8.5,15.5) {0};
\fill[paritycolor] (9,15) rectangle (10,16);
\node at (9.5,15.5) {1};
\fill[paritycolor] (10,15) rectangle (11,16);
\node at (10.5,15.5) {1};
\fill[paritycolor] (11,15) rectangle (12,16);
\node at (11.5,15.5) {0};
\fill[paritycolor] (12,15) rectangle (13,16);
\node at (12.5,15.5) {1};
\fill[paritycolor] (13,15) rectangle (14,16);
\node at (13.5,15.5) {1};
\fill[paritycolor] (14,15) rectangle (15,16);
\node at (14.5,15.5) {1};
\fill[paritycolor] (15,15) rectangle (16,16);
\node at (15.5,15.5) {1};
\draw[step=1,gray!60,thin] (0,0) grid (16,16);
\draw[thick] (0,0) rectangle (16,16);
\node at (0.5,-.48) {$ 1 $};
\node at (1.5,-.48) {$ 2 $};
\node at (2.5,-.48) {$ 3 $};
\node at (3.5,-.48) {$ 4 $};
\node at (4.5,-.48) {$ 5 $};
\node at (5.5,-.48) {$ 6 $};
\node at (6.5,-.48) {$ 7 $};
\node at (7.5,-.48) {$ 8 $};
\node at (8.5,-.48) {$ 9 $};
\node at (9.5,-.48) {$ 10 $};
\node at (10.5,-.48) {$ 11 $};
\node at (11.5,-.48) {$ 12 $};
\node at (12.5,-.48) {$ 13 $};
\node at (13.5,-.48) {$ 14 $};
\node at (14.5,-.48) {$ 15 $};
\node at (15.5,-.48) {$ 16 $};
\node at (-.5,0.5) {$ 1 $};
\node at (-.5,1.5) {$ 2 $};
\node at (-.5,2.5) {$ 3 $};
\node at (-.5,3.5) {$ 4 $};
\node at (-.5,4.5) {$ 5 $};
\node at (-.5,5.5) {$ 6 $};
\node at (-.5,6.5) {$ 7 $};
\node at (-.5,7.5) {$ 8 $};
\node at (-.5,8.5) {$ 9 $};
\node at (-.5,9.5) {$ 10 $};
\node at (-.5,10.5) {$ 11 $};
\node at (-.5,11.5) {$ 12 $};
\node at (-.5,12.5) {$ 13 $};
\node at (-.5,13.5) {$ 14 $};
\node at (-.5,14.5) {$ 15 $};
\node at (-.5,15.5) {$ 16 $};
\draw[thick] (4,0)--(4,4)--(0,4);
\draw[thick] (6,0)--(6,6)--(0,6);
\draw (0,-1)--(0,-1.2)--(4,-1.2)--(4,-1);
\node at (2.0,-1.65) {$ 4 $};
\draw (4,-1)--(4,-1.2)--(6,-1.2)--(6,-1);
\node at (5.0,-1.65) {$ 2 $};
\draw (6,-1)--(6,-1.2)--(16,-1.2)--(16,-1);
\node at (11.0,-1.65) {$ 10 $};
\draw (-1,0)--(-1.2,0)--(-1.2,4)--(-1,4);
\node at (-1.65,2.0) {$ 4 $};
\draw (-1,4)--(-1.2,4)--(-1.2,6)--(-1,6);
\node at (-1.65,5.0) {$ 2 $};
\draw (-1,6)--(-1.2,6)--(-1.2,16)--(-1,16);
\node at (-1.65,11.0) {$ 10 $};
\draw[red!75!black,line width=.8pt] (0,2) rectangle (16,3);
\draw[red!75!black,line width=.8pt] (9,0) rectangle (10,16);
\end{tikzpicture}%
}
\par\smallskip\footnotesize (a) Original array $\bm X$
\end{minipage}\hfill%
\begin{minipage}[t]{.32\linewidth}\centering\vspace{0pt}
\resizebox{\linewidth}{!}{%
\begin{tikzpicture}[x=.30cm,y=-.30cm,font=\scriptsize]
\tikzset{every node/.style={inner sep=0pt}}
\path[use as bounding box] (-2.2,-2.2) rectangle (16.3,16.3);
\fill[seedcolor] (0,0) rectangle (1,1);
\node at (0.5,0.5) {0};
\fill[seedcolor] (1,0) rectangle (2,1);
\node at (1.5,0.5) {1};
\fill[seedcolor] (2,0) rectangle (3,1);
\node at (2.5,0.5) {1};
\fill[seedcolor] (3,0) rectangle (4,1);
\node at (3.5,0.5) {0};
\fill[innercolor] (4,0) rectangle (5,1);
\node at (4.5,0.5) {1};
\fill[innercolor] (5,0) rectangle (6,1);
\node at (5.5,0.5) {0};
\fill[outercolor] (6,0) rectangle (7,1);
\node at (6.5,0.5) {1};
\fill[outercolor] (7,0) rectangle (8,1);
\node at (7.5,0.5) {0};
\fill[outercolor] (8,0) rectangle (9,1);
\node at (8.5,0.5) {1};
\fill[outercolor] (9,0) rectangle (10,1);
\node at (9.5,0.5) {1};
\fill[outercolor] (10,0) rectangle (11,1);
\node at (10.5,0.5) {0};
\fill[outercolor] (11,0) rectangle (12,1);
\node at (11.5,0.5) {1};
\fill[outercolor] (12,0) rectangle (13,1);
\node at (12.5,0.5) {1};
\fill[outercolor] (13,0) rectangle (14,1);
\node at (13.5,0.5) {0};
\fill[outercolor] (14,0) rectangle (15,1);
\node at (14.5,0.5) {0};
\fill[seedcolor] (0,1) rectangle (1,2);
\node at (0.5,1.5) {0};
\fill[seedcolor] (1,1) rectangle (2,2);
\node at (1.5,1.5) {1};
\fill[seedcolor] (2,1) rectangle (3,2);
\node at (2.5,1.5) {0};
\fill[seedcolor] (3,1) rectangle (4,2);
\node at (3.5,1.5) {0};
\fill[innercolor] (4,1) rectangle (5,2);
\node at (4.5,1.5) {1};
\fill[innercolor] (5,1) rectangle (6,2);
\node at (5.5,1.5) {0};
\fill[outercolor] (6,1) rectangle (7,2);
\node at (6.5,1.5) {1};
\fill[outercolor] (7,1) rectangle (8,2);
\node at (7.5,1.5) {0};
\fill[outercolor] (8,1) rectangle (9,2);
\node at (8.5,1.5) {0};
\fill[outercolor] (9,1) rectangle (10,2);
\node at (9.5,1.5) {1};
\fill[outercolor] (10,1) rectangle (11,2);
\node at (10.5,1.5) {0};
\fill[outercolor] (11,1) rectangle (12,2);
\node at (11.5,1.5) {1};
\fill[outercolor] (12,1) rectangle (13,2);
\node at (12.5,1.5) {0};
\fill[outercolor] (13,1) rectangle (14,2);
\node at (13.5,1.5) {1};
\fill[outercolor] (14,1) rectangle (15,2);
\node at (14.5,1.5) {0};
\fill[seedcolor] (0,2) rectangle (1,3);
\node at (0.5,2.5) {0};
\fill[seedcolor] (1,2) rectangle (2,3);
\node at (1.5,2.5) {0};
\fill[seedcolor] (2,2) rectangle (3,3);
\node at (2.5,2.5) {0};
\fill[seedcolor] (3,2) rectangle (4,3);
\node at (3.5,2.5) {0};
\fill[innercolor] (4,2) rectangle (5,3);
\node at (4.5,2.5) {1};
\fill[innercolor] (5,2) rectangle (6,3);
\node at (5.5,2.5) {1};
\fill[outercolor] (6,2) rectangle (7,3);
\node at (6.5,2.5) {1};
\fill[outercolor] (7,2) rectangle (8,3);
\node at (7.5,2.5) {1};
\fill[outercolor] (8,2) rectangle (9,3);
\node at (8.5,2.5) {1};
\fill[outercolor] (9,2) rectangle (10,3);
\node at (9.5,2.5) {1};
\fill[outercolor] (10,2) rectangle (11,3);
\node at (10.5,2.5) {0};
\fill[outercolor] (11,2) rectangle (12,3);
\node at (11.5,2.5) {1};
\fill[outercolor] (12,2) rectangle (13,3);
\node at (12.5,2.5) {0};
\fill[outercolor] (13,2) rectangle (14,3);
\node at (13.5,2.5) {1};
\fill[outercolor] (14,2) rectangle (15,3);
\node at (14.5,2.5) {1};
\fill[innercolor] (0,3) rectangle (1,4);
\node at (0.5,3.5) {1};
\fill[innercolor] (1,3) rectangle (2,4);
\node at (1.5,3.5) {1};
\fill[innercolor] (2,3) rectangle (3,4);
\node at (2.5,3.5) {0};
\fill[innercolor] (3,3) rectangle (4,4);
\node at (3.5,3.5) {1};
\node at (4.5,3.5) {1};
\node at (5.5,3.5) {0};
\fill[outercolor] (6,3) rectangle (7,4);
\node at (6.5,3.5) {0};
\fill[outercolor] (7,3) rectangle (8,4);
\node at (7.5,3.5) {1};
\fill[outercolor] (8,3) rectangle (9,4);
\node at (8.5,3.5) {1};
\fill[outercolor] (9,3) rectangle (10,4);
\node at (9.5,3.5) {1};
\fill[outercolor] (10,3) rectangle (11,4);
\node at (10.5,3.5) {1};
\fill[outercolor] (11,3) rectangle (12,4);
\node at (11.5,3.5) {1};
\fill[outercolor] (12,3) rectangle (13,4);
\node at (12.5,3.5) {0};
\fill[outercolor] (13,3) rectangle (14,4);
\node at (13.5,3.5) {0};
\fill[outercolor] (14,3) rectangle (15,4);
\node at (14.5,3.5) {1};
\fill[innercolor] (0,4) rectangle (1,5);
\node at (0.5,4.5) {0};
\fill[innercolor] (1,4) rectangle (2,5);
\node at (1.5,4.5) {0};
\fill[innercolor] (2,4) rectangle (3,5);
\node at (2.5,4.5) {1};
\fill[innercolor] (3,4) rectangle (4,5);
\node at (3.5,4.5) {0};
\node at (4.5,4.5) {1};
\node at (5.5,4.5) {0};
\fill[outercolor] (6,4) rectangle (7,5);
\node at (6.5,4.5) {1};
\fill[outercolor] (7,4) rectangle (8,5);
\node at (7.5,4.5) {1};
\fill[outercolor] (8,4) rectangle (9,5);
\node at (8.5,4.5) {0};
\fill[outercolor] (9,4) rectangle (10,5);
\node at (9.5,4.5) {0};
\fill[outercolor] (10,4) rectangle (11,5);
\node at (10.5,4.5) {0};
\fill[outercolor] (11,4) rectangle (12,5);
\node at (11.5,4.5) {1};
\fill[outercolor] (12,4) rectangle (13,5);
\node at (12.5,4.5) {1};
\fill[outercolor] (13,4) rectangle (14,5);
\node at (13.5,4.5) {1};
\fill[outercolor] (14,4) rectangle (15,5);
\node at (14.5,4.5) {0};
\fill[outercolor] (0,5) rectangle (1,6);
\node at (0.5,5.5) {1};
\fill[outercolor] (1,5) rectangle (2,6);
\node at (1.5,5.5) {1};
\fill[outercolor] (2,5) rectangle (3,6);
\node at (2.5,5.5) {0};
\fill[outercolor] (3,5) rectangle (4,6);
\node at (3.5,5.5) {1};
\fill[outercolor] (4,5) rectangle (5,6);
\node at (4.5,5.5) {0};
\fill[outercolor] (5,5) rectangle (6,6);
\node at (5.5,5.5) {0};
\node at (6.5,5.5) {1};
\node at (7.5,5.5) {0};
\node at (8.5,5.5) {1};
\node at (9.5,5.5) {1};
\node at (10.5,5.5) {0};
\node at (11.5,5.5) {1};
\node at (12.5,5.5) {1};
\node at (13.5,5.5) {0};
\fill[paritycolor] (14,5) rectangle (15,6);
\node at (14.5,5.5) {0};
\fill[outercolor] (0,6) rectangle (1,7);
\node at (0.5,6.5) {0};
\fill[outercolor] (1,6) rectangle (2,7);
\node at (1.5,6.5) {0};
\fill[outercolor] (2,6) rectangle (3,7);
\node at (2.5,6.5) {0};
\fill[outercolor] (3,6) rectangle (4,7);
\node at (3.5,6.5) {0};
\fill[outercolor] (4,6) rectangle (5,7);
\node at (4.5,6.5) {0};
\fill[outercolor] (5,6) rectangle (6,7);
\node at (5.5,6.5) {0};
\node at (6.5,6.5) {0};
\node at (7.5,6.5) {0};
\node at (8.5,6.5) {1};
\node at (9.5,6.5) {1};
\node at (10.5,6.5) {1};
\node at (11.5,6.5) {1};
\node at (12.5,6.5) {1};
\node at (13.5,6.5) {0};
\fill[paritycolor] (14,6) rectangle (15,7);
\node at (14.5,6.5) {1};
\fill[outercolor] (0,7) rectangle (1,8);
\node at (0.5,7.5) {1};
\fill[outercolor] (1,7) rectangle (2,8);
\node at (1.5,7.5) {0};
\fill[outercolor] (2,7) rectangle (3,8);
\node at (2.5,7.5) {0};
\fill[outercolor] (3,7) rectangle (4,8);
\node at (3.5,7.5) {1};
\fill[outercolor] (4,7) rectangle (5,8);
\node at (4.5,7.5) {0};
\fill[outercolor] (5,7) rectangle (6,8);
\node at (5.5,7.5) {0};
\node at (6.5,7.5) {0};
\node at (7.5,7.5) {0};
\node at (8.5,7.5) {0};
\node at (9.5,7.5) {1};
\node at (10.5,7.5) {1};
\node at (11.5,7.5) {0};
\node at (12.5,7.5) {0};
\node at (13.5,7.5) {1};
\fill[paritycolor] (14,7) rectangle (15,8);
\node at (14.5,7.5) {1};
\fill[outercolor] (0,8) rectangle (1,9);
\node at (0.5,8.5) {0};
\fill[outercolor] (1,8) rectangle (2,9);
\node at (1.5,8.5) {1};
\fill[outercolor] (2,8) rectangle (3,9);
\node at (2.5,8.5) {1};
\fill[outercolor] (3,8) rectangle (4,9);
\node at (3.5,8.5) {0};
\fill[outercolor] (4,8) rectangle (5,9);
\node at (4.5,8.5) {1};
\fill[outercolor] (5,8) rectangle (6,9);
\node at (5.5,8.5) {0};
\node at (6.5,8.5) {1};
\node at (7.5,8.5) {1};
\node at (8.5,8.5) {0};
\node at (9.5,8.5) {0};
\node at (10.5,8.5) {0};
\node at (11.5,8.5) {1};
\node at (12.5,8.5) {0};
\node at (13.5,8.5) {1};
\fill[paritycolor] (14,8) rectangle (15,9);
\node at (14.5,8.5) {0};
\fill[outercolor] (0,9) rectangle (1,10);
\node at (0.5,9.5) {0};
\fill[outercolor] (1,9) rectangle (2,10);
\node at (1.5,9.5) {1};
\fill[outercolor] (2,9) rectangle (3,10);
\node at (2.5,9.5) {0};
\fill[outercolor] (3,9) rectangle (4,10);
\node at (3.5,9.5) {1};
\fill[outercolor] (4,9) rectangle (5,10);
\node at (4.5,9.5) {0};
\fill[outercolor] (5,9) rectangle (6,10);
\node at (5.5,9.5) {0};
\node at (6.5,9.5) {0};
\node at (7.5,9.5) {1};
\node at (8.5,9.5) {0};
\node at (9.5,9.5) {1};
\node at (10.5,9.5) {0};
\node at (11.5,9.5) {0};
\node at (12.5,9.5) {1};
\node at (13.5,9.5) {1};
\fill[paritycolor] (14,9) rectangle (15,10);
\node at (14.5,9.5) {1};
\fill[outercolor] (0,10) rectangle (1,11);
\node at (0.5,10.5) {0};
\fill[outercolor] (1,10) rectangle (2,11);
\node at (1.5,10.5) {1};
\fill[outercolor] (2,10) rectangle (3,11);
\node at (2.5,10.5) {0};
\fill[outercolor] (3,10) rectangle (4,11);
\node at (3.5,10.5) {1};
\fill[outercolor] (4,10) rectangle (5,11);
\node at (4.5,10.5) {0};
\fill[outercolor] (5,10) rectangle (6,11);
\node at (5.5,10.5) {1};
\node at (6.5,10.5) {1};
\node at (7.5,10.5) {0};
\node at (8.5,10.5) {1};
\node at (9.5,10.5) {0};
\node at (10.5,10.5) {1};
\node at (11.5,10.5) {0};
\node at (12.5,10.5) {1};
\node at (13.5,10.5) {0};
\fill[paritycolor] (14,10) rectangle (15,11);
\node at (14.5,10.5) {1};
\fill[outercolor] (0,11) rectangle (1,12);
\node at (0.5,11.5) {1};
\fill[outercolor] (1,11) rectangle (2,12);
\node at (1.5,11.5) {1};
\fill[outercolor] (2,11) rectangle (3,12);
\node at (2.5,11.5) {0};
\fill[outercolor] (3,11) rectangle (4,12);
\node at (3.5,11.5) {0};
\fill[outercolor] (4,11) rectangle (5,12);
\node at (4.5,11.5) {0};
\fill[outercolor] (5,11) rectangle (6,12);
\node at (5.5,11.5) {0};
\node at (6.5,11.5) {0};
\node at (7.5,11.5) {1};
\node at (8.5,11.5) {1};
\node at (9.5,11.5) {0};
\node at (10.5,11.5) {1};
\node at (11.5,11.5) {0};
\node at (12.5,11.5) {0};
\node at (13.5,11.5) {0};
\fill[paritycolor] (14,11) rectangle (15,12);
\node at (14.5,11.5) {0};
\fill[outercolor] (0,12) rectangle (1,13);
\node at (0.5,12.5) {1};
\fill[outercolor] (1,12) rectangle (2,13);
\node at (1.5,12.5) {1};
\fill[outercolor] (2,12) rectangle (3,13);
\node at (2.5,12.5) {1};
\fill[outercolor] (3,12) rectangle (4,13);
\node at (3.5,12.5) {1};
\fill[outercolor] (4,12) rectangle (5,13);
\node at (4.5,12.5) {1};
\fill[outercolor] (5,12) rectangle (6,13);
\node at (5.5,12.5) {1};
\node at (6.5,12.5) {1};
\node at (7.5,12.5) {1};
\node at (8.5,12.5) {0};
\node at (9.5,12.5) {1};
\node at (10.5,12.5) {0};
\node at (11.5,12.5) {1};
\node at (12.5,12.5) {0};
\node at (13.5,12.5) {1};
\fill[paritycolor] (14,12) rectangle (15,13);
\node at (14.5,12.5) {1};
\fill[outercolor] (0,13) rectangle (1,14);
\node at (0.5,13.5) {1};
\fill[outercolor] (1,13) rectangle (2,14);
\node at (1.5,13.5) {0};
\fill[outercolor] (2,13) rectangle (3,14);
\node at (2.5,13.5) {0};
\fill[outercolor] (3,13) rectangle (4,14);
\node at (3.5,13.5) {1};
\fill[outercolor] (4,13) rectangle (5,14);
\node at (4.5,13.5) {0};
\fill[outercolor] (5,13) rectangle (6,14);
\node at (5.5,13.5) {1};
\node at (6.5,13.5) {0};
\node at (7.5,13.5) {0};
\node at (8.5,13.5) {1};
\node at (9.5,13.5) {0};
\node at (10.5,13.5) {0};
\node at (11.5,13.5) {1};
\node at (12.5,13.5) {1};
\node at (13.5,13.5) {1};
\fill[paritycolor] (14,13) rectangle (15,14);
\node at (14.5,13.5) {0};
\fill[outercolor] (0,14) rectangle (1,15);
\node at (0.5,14.5) {1};
\fill[outercolor] (1,14) rectangle (2,15);
\node at (1.5,14.5) {0};
\fill[outercolor] (2,14) rectangle (3,15);
\node at (2.5,14.5) {1};
\fill[outercolor] (3,14) rectangle (4,15);
\node at (3.5,14.5) {1};
\fill[outercolor] (4,14) rectangle (5,15);
\node at (4.5,14.5) {1};
\fill[outercolor] (5,14) rectangle (6,15);
\node at (5.5,14.5) {0};
\fill[paritycolor] (6,14) rectangle (7,15);
\node at (6.5,14.5) {0};
\fill[paritycolor] (7,14) rectangle (8,15);
\node at (7.5,14.5) {0};
\fill[paritycolor] (8,14) rectangle (9,15);
\node at (8.5,14.5) {0};
\fill[paritycolor] (9,14) rectangle (10,15);
\node at (9.5,14.5) {1};
\fill[paritycolor] (10,14) rectangle (11,15);
\node at (10.5,14.5) {0};
\fill[paritycolor] (11,14) rectangle (12,15);
\node at (11.5,14.5) {1};
\fill[paritycolor] (12,14) rectangle (13,15);
\node at (12.5,14.5) {1};
\fill[paritycolor] (13,14) rectangle (14,15);
\node at (13.5,14.5) {1};
\fill[paritycolor] (14,14) rectangle (15,15);
\node at (14.5,14.5) {1};
\draw[step=1,gray!60,thin] (0,0) grid (15,15);
\draw[thick] (0,0) rectangle (15,15);
\node at (0.5,-.48) {$ 1 $};
\node at (1.5,-.48) {$ 2 $};
\node at (2.5,-.48) {$ 3 $};
\node at (3.5,-.48) {$ 4 $};
\node at (4.5,-.48) {$ 5 $};
\node at (5.5,-.48) {$ 6 $};
\node at (6.5,-.48) {$ 7 $};
\node at (7.5,-.48) {$ 8 $};
\node at (8.5,-.48) {$ 9 $};
\node at (9.5,-.48) {$ 10 $};
\node at (10.5,-.48) {$ 11 $};
\node at (11.5,-.48) {$ 12 $};
\node at (12.5,-.48) {$ 13 $};
\node at (13.5,-.48) {$ 14 $};
\node at (14.5,-.48) {$ 15 $};
\node at (-.5,0.5) {$ 1 $};
\node at (-.5,1.5) {$ 2 $};
\node at (-.5,2.5) {$ 3 $};
\node at (-.5,3.5) {$ 4 $};
\node at (-.5,4.5) {$ 5 $};
\node at (-.5,5.5) {$ 6 $};
\node at (-.5,6.5) {$ 7 $};
\node at (-.5,7.5) {$ 8 $};
\node at (-.5,8.5) {$ 9 $};
\node at (-.5,9.5) {$ 10 $};
\node at (-.5,10.5) {$ 11 $};
\node at (-.5,11.5) {$ 12 $};
\node at (-.5,12.5) {$ 13 $};
\node at (-.5,13.5) {$ 14 $};
\node at (-.5,14.5) {$ 15 $};
\draw[thick] (4,0)--(4,3)--(0,3);
\draw[thick] (6,0)--(6,5)--(0,5);
\draw (0,-1)--(0,-1.2)--(4,-1.2)--(4,-1);
\node at (2.0,-1.65) {$ 4 $};
\draw (4,-1)--(4,-1.2)--(6,-1.2)--(6,-1);
\node at (5.0,-1.65) {$ 2 $};
\draw (6,-1)--(6,-1.2)--(15,-1.2)--(15,-1);
\node at (10.5,-1.65) {$ 9 $};
\draw (-1,0)--(-1.2,0)--(-1.2,3)--(-1,3);
\node at (-1.65,1.5) {$ 3 $};
\draw (-1,3)--(-1.2,3)--(-1.2,5)--(-1,5);
\node at (-1.65,4.0) {$ 2 $};
\draw (-1,5)--(-1.2,5)--(-1.2,15)--(-1,15);
\node at (-1.65,10.0) {$ 10 $};
\end{tikzpicture}%
}
\par\smallskip\footnotesize (b) Deletion output $\bm Y$
\end{minipage}\hfill%
\begin{minipage}[t]{.32\linewidth}\centering\vspace{0pt}
\resizebox{\linewidth}{!}{%
\begin{tikzpicture}[x=.30cm,y=-.30cm,font=\scriptsize]
\tikzset{every node/.style={inner sep=0pt}}
\path[use as bounding box] (-2.2,-2.2) rectangle (16.3,16.3);
\fill[seedcolor] (0,0) rectangle (1,1);
\node at (0.5,0.5) {0};
\fill[seedcolor] (1,0) rectangle (2,1);
\node at (1.5,0.5) {1};
\fill[seedcolor] (2,0) rectangle (3,1);
\node at (2.5,0.5) {1};
\fill[seedcolor] (3,0) rectangle (4,1);
\node at (3.5,0.5) {0};
\fill[innercolor] (4,0) rectangle (5,1);
\node at (4.5,0.5) {1};
\fill[innercolor] (5,0) rectangle (6,1);
\node at (5.5,0.5) {0};
\fill[outercolor] (6,0) rectangle (7,1);
\node at (6.5,0.5) {1};
\fill[outercolor] (7,0) rectangle (8,1);
\node at (7.5,0.5) {0};
\fill[outercolor] (8,0) rectangle (9,1);
\node at (8.5,0.5) {1};
\fill[outercolor] (9,0) rectangle (10,1);
\node at (9.5,0.5) {1};
\fill[outercolor] (10,0) rectangle (11,1);
\node at (10.5,0.5) {0};
\fill[outercolor] (11,0) rectangle (12,1);
\node at (11.5,0.5) {1};
\fill[outercolor] (12,0) rectangle (13,1);
\node at (12.5,0.5) {1};
\fill[outercolor] (13,0) rectangle (14,1);
\node at (13.5,0.5) {0};
\fill[outercolor] (14,0) rectangle (15,1);
\node at (14.5,0.5) {0};
\fill[outercolor] (15,0) rectangle (16,1);
\node at (15.5,0.5) {0};
\fill[seedcolor] (0,1) rectangle (1,2);
\node at (0.5,1.5) {0};
\fill[seedcolor] (1,1) rectangle (2,2);
\node at (1.5,1.5) {1};
\fill[seedcolor] (2,1) rectangle (3,2);
\node at (2.5,1.5) {0};
\fill[seedcolor] (3,1) rectangle (4,2);
\node at (3.5,1.5) {0};
\fill[innercolor] (4,1) rectangle (5,2);
\node at (4.5,1.5) {1};
\fill[innercolor] (5,1) rectangle (6,2);
\node at (5.5,1.5) {0};
\fill[outercolor] (6,1) rectangle (7,2);
\node at (6.5,1.5) {1};
\fill[outercolor] (7,1) rectangle (8,2);
\node at (7.5,1.5) {0};
\fill[outercolor] (8,1) rectangle (9,2);
\node at (8.5,1.5) {0};
\fill[outercolor] (9,1) rectangle (10,2);
\node at (9.5,1.5) {1};
\fill[outercolor] (10,1) rectangle (11,2);
\node at (10.5,1.5) {0};
\fill[outercolor] (11,1) rectangle (12,2);
\node at (11.5,1.5) {1};
\fill[outercolor] (12,1) rectangle (13,2);
\node at (12.5,1.5) {0};
\fill[outercolor] (13,1) rectangle (14,2);
\node at (13.5,1.5) {1};
\fill[outercolor] (14,1) rectangle (15,2);
\node at (14.5,1.5) {0};
\fill[outercolor] (15,1) rectangle (16,2);
\node at (15.5,1.5) {0};
\fill[seedcolor] (0,2) rectangle (1,3);
\node at (0.5,2.5) {0};
\fill[seedcolor] (1,2) rectangle (2,3);
\node at (1.5,2.5) {0};
\fill[seedcolor] (2,2) rectangle (3,3);
\node at (2.5,2.5) {0};
\fill[seedcolor] (3,2) rectangle (4,3);
\node at (3.5,2.5) {0};
\fill[innercolor] (4,2) rectangle (5,3);
\node at (4.5,2.5) {1};
\fill[innercolor] (5,2) rectangle (6,3);
\node at (5.5,2.5) {1};
\fill[outercolor] (6,2) rectangle (7,3);
\node at (6.5,2.5) {1};
\fill[outercolor] (7,2) rectangle (8,3);
\node at (7.5,2.5) {1};
\fill[outercolor] (8,2) rectangle (9,3);
\node at (8.5,2.5) {1};
\fill[outercolor] (9,2) rectangle (10,3);
\node at (9.5,2.5) {1};
\fill[outercolor] (10,2) rectangle (11,3);
\node at (10.5,2.5) {0};
\fill[outercolor] (11,2) rectangle (12,3);
\node at (11.5,2.5) {1};
\fill[outercolor] (12,2) rectangle (13,3);
\node at (12.5,2.5) {0};
\fill[outercolor] (13,2) rectangle (14,3);
\node at (13.5,2.5) {1};
\fill[outercolor] (14,2) rectangle (15,3);
\node at (14.5,2.5) {1};
\fill[outercolor] (15,2) rectangle (16,3);
\node at (15.5,2.5) {1};
\fill[seedcolor] (0,3) rectangle (1,4);
\node at (0.5,3.5) {1};
\fill[seedcolor] (1,3) rectangle (2,4);
\node at (1.5,3.5) {1};
\fill[seedcolor] (2,3) rectangle (3,4);
\node at (2.5,3.5) {1};
\fill[seedcolor] (3,3) rectangle (4,4);
\node at (3.5,3.5) {0};
\fill[innercolor] (4,3) rectangle (5,4);
\node at (4.5,3.5) {0};
\fill[innercolor] (5,3) rectangle (6,4);
\node at (5.5,3.5) {0};
\fill[outercolor] (6,3) rectangle (7,4);
\node at (6.5,3.5) {0};
\fill[outercolor] (7,3) rectangle (8,4);
\node at (7.5,3.5) {1};
\fill[outercolor] (8,3) rectangle (9,4);
\node at (8.5,3.5) {0};
\fill[outercolor] (9,3) rectangle (10,4);
\node at (9.5,3.5) {0};
\fill[outercolor] (10,3) rectangle (11,4);
\node at (10.5,3.5) {1};
\fill[outercolor] (11,3) rectangle (12,4);
\node at (11.5,3.5) {1};
\fill[outercolor] (12,3) rectangle (13,4);
\node at (12.5,3.5) {0};
\fill[outercolor] (13,3) rectangle (14,4);
\node at (13.5,3.5) {1};
\fill[outercolor] (14,3) rectangle (15,4);
\node at (14.5,3.5) {0};
\fill[outercolor] (15,3) rectangle (16,4);
\node at (15.5,3.5) {1};
\fill[innercolor] (0,4) rectangle (1,5);
\node at (0.5,4.5) {1};
\fill[innercolor] (1,4) rectangle (2,5);
\node at (1.5,4.5) {1};
\fill[innercolor] (2,4) rectangle (3,5);
\node at (2.5,4.5) {0};
\fill[innercolor] (3,4) rectangle (4,5);
\node at (3.5,4.5) {1};
\node at (4.5,4.5) {1};
\node at (5.5,4.5) {0};
\fill[outercolor] (6,4) rectangle (7,5);
\node at (6.5,4.5) {0};
\fill[outercolor] (7,4) rectangle (8,5);
\node at (7.5,4.5) {1};
\fill[outercolor] (8,4) rectangle (9,5);
\node at (8.5,4.5) {1};
\fill[outercolor] (9,4) rectangle (10,5);
\node at (9.5,4.5) {1};
\fill[outercolor] (10,4) rectangle (11,5);
\node at (10.5,4.5) {1};
\fill[outercolor] (11,4) rectangle (12,5);
\node at (11.5,4.5) {1};
\fill[outercolor] (12,4) rectangle (13,5);
\node at (12.5,4.5) {0};
\fill[outercolor] (13,4) rectangle (14,5);
\node at (13.5,4.5) {0};
\fill[outercolor] (14,4) rectangle (15,5);
\node at (14.5,4.5) {1};
\fill[outercolor] (15,4) rectangle (16,5);
\node at (15.5,4.5) {0};
\fill[innercolor] (0,5) rectangle (1,6);
\node at (0.5,5.5) {0};
\fill[innercolor] (1,5) rectangle (2,6);
\node at (1.5,5.5) {0};
\fill[innercolor] (2,5) rectangle (3,6);
\node at (2.5,5.5) {1};
\fill[innercolor] (3,5) rectangle (4,6);
\node at (3.5,5.5) {0};
\node at (4.5,5.5) {1};
\node at (5.5,5.5) {0};
\fill[outercolor] (6,5) rectangle (7,6);
\node at (6.5,5.5) {1};
\fill[outercolor] (7,5) rectangle (8,6);
\node at (7.5,5.5) {1};
\fill[outercolor] (8,5) rectangle (9,6);
\node at (8.5,5.5) {0};
\fill[outercolor] (9,5) rectangle (10,6);
\node at (9.5,5.5) {0};
\fill[outercolor] (10,5) rectangle (11,6);
\node at (10.5,5.5) {0};
\fill[outercolor] (11,5) rectangle (12,6);
\node at (11.5,5.5) {1};
\fill[outercolor] (12,5) rectangle (13,6);
\node at (12.5,5.5) {1};
\fill[outercolor] (13,5) rectangle (14,6);
\node at (13.5,5.5) {1};
\fill[outercolor] (14,5) rectangle (15,6);
\node at (14.5,5.5) {0};
\fill[outercolor] (15,5) rectangle (16,6);
\node at (15.5,5.5) {1};
\fill[outercolor] (0,6) rectangle (1,7);
\node at (0.5,6.5) {1};
\fill[outercolor] (1,6) rectangle (2,7);
\node at (1.5,6.5) {1};
\fill[outercolor] (2,6) rectangle (3,7);
\node at (2.5,6.5) {0};
\fill[outercolor] (3,6) rectangle (4,7);
\node at (3.5,6.5) {1};
\fill[outercolor] (4,6) rectangle (5,7);
\node at (4.5,6.5) {0};
\fill[outercolor] (5,6) rectangle (6,7);
\node at (5.5,6.5) {0};
\node at (6.5,6.5) {1};
\node at (7.5,6.5) {0};
\node at (8.5,6.5) {1};
\node at (9.5,6.5) {1};
\node at (10.5,6.5) {0};
\node at (11.5,6.5) {1};
\node at (12.5,6.5) {1};
\node at (13.5,6.5) {0};
\fill[paritycolor] (14,6) rectangle (15,7);
\node at (14.5,6.5) {0};
\node at (15.5,6.5) {0};
\fill[outercolor] (0,7) rectangle (1,8);
\node at (0.5,7.5) {0};
\fill[outercolor] (1,7) rectangle (2,8);
\node at (1.5,7.5) {0};
\fill[outercolor] (2,7) rectangle (3,8);
\node at (2.5,7.5) {0};
\fill[outercolor] (3,7) rectangle (4,8);
\node at (3.5,7.5) {0};
\fill[outercolor] (4,7) rectangle (5,8);
\node at (4.5,7.5) {0};
\fill[outercolor] (5,7) rectangle (6,8);
\node at (5.5,7.5) {0};
\node at (6.5,7.5) {0};
\node at (7.5,7.5) {0};
\node at (8.5,7.5) {1};
\node at (9.5,7.5) {1};
\node at (10.5,7.5) {1};
\node at (11.5,7.5) {1};
\node at (12.5,7.5) {1};
\node at (13.5,7.5) {0};
\fill[paritycolor] (14,7) rectangle (15,8);
\node at (14.5,7.5) {1};
\node at (15.5,7.5) {0};
\fill[outercolor] (0,8) rectangle (1,9);
\node at (0.5,8.5) {1};
\fill[outercolor] (1,8) rectangle (2,9);
\node at (1.5,8.5) {0};
\fill[outercolor] (2,8) rectangle (3,9);
\node at (2.5,8.5) {0};
\fill[outercolor] (3,8) rectangle (4,9);
\node at (3.5,8.5) {1};
\fill[outercolor] (4,8) rectangle (5,9);
\node at (4.5,8.5) {0};
\fill[outercolor] (5,8) rectangle (6,9);
\node at (5.5,8.5) {0};
\node at (6.5,8.5) {0};
\node at (7.5,8.5) {0};
\node at (8.5,8.5) {0};
\node at (9.5,8.5) {1};
\node at (10.5,8.5) {1};
\node at (11.5,8.5) {0};
\node at (12.5,8.5) {0};
\node at (13.5,8.5) {1};
\fill[paritycolor] (14,8) rectangle (15,9);
\node at (14.5,8.5) {1};
\node at (15.5,8.5) {0};
\fill[outercolor] (0,9) rectangle (1,10);
\node at (0.5,9.5) {0};
\fill[outercolor] (1,9) rectangle (2,10);
\node at (1.5,9.5) {1};
\fill[outercolor] (2,9) rectangle (3,10);
\node at (2.5,9.5) {1};
\fill[outercolor] (3,9) rectangle (4,10);
\node at (3.5,9.5) {0};
\fill[outercolor] (4,9) rectangle (5,10);
\node at (4.5,9.5) {1};
\fill[outercolor] (5,9) rectangle (6,10);
\node at (5.5,9.5) {0};
\node at (6.5,9.5) {1};
\node at (7.5,9.5) {1};
\node at (8.5,9.5) {0};
\node at (9.5,9.5) {0};
\node at (10.5,9.5) {0};
\node at (11.5,9.5) {1};
\node at (12.5,9.5) {0};
\node at (13.5,9.5) {1};
\fill[paritycolor] (14,9) rectangle (15,10);
\node at (14.5,9.5) {0};
\node at (15.5,9.5) {1};
\fill[outercolor] (0,10) rectangle (1,11);
\node at (0.5,10.5) {0};
\fill[outercolor] (1,10) rectangle (2,11);
\node at (1.5,10.5) {1};
\fill[outercolor] (2,10) rectangle (3,11);
\node at (2.5,10.5) {0};
\fill[outercolor] (3,10) rectangle (4,11);
\node at (3.5,10.5) {1};
\fill[outercolor] (4,10) rectangle (5,11);
\node at (4.5,10.5) {0};
\fill[outercolor] (5,10) rectangle (6,11);
\node at (5.5,10.5) {0};
\node at (6.5,10.5) {0};
\node at (7.5,10.5) {1};
\node at (8.5,10.5) {0};
\node at (9.5,10.5) {1};
\node at (10.5,10.5) {0};
\node at (11.5,10.5) {0};
\node at (12.5,10.5) {1};
\node at (13.5,10.5) {1};
\fill[paritycolor] (14,10) rectangle (15,11);
\node at (14.5,10.5) {1};
\node at (15.5,10.5) {1};
\fill[outercolor] (0,11) rectangle (1,12);
\node at (0.5,11.5) {0};
\fill[outercolor] (1,11) rectangle (2,12);
\node at (1.5,11.5) {1};
\fill[outercolor] (2,11) rectangle (3,12);
\node at (2.5,11.5) {0};
\fill[outercolor] (3,11) rectangle (4,12);
\node at (3.5,11.5) {1};
\fill[outercolor] (4,11) rectangle (5,12);
\node at (4.5,11.5) {0};
\fill[outercolor] (5,11) rectangle (6,12);
\node at (5.5,11.5) {1};
\node at (6.5,11.5) {1};
\node at (7.5,11.5) {0};
\node at (8.5,11.5) {1};
\node at (9.5,11.5) {0};
\node at (10.5,11.5) {1};
\node at (11.5,11.5) {0};
\node at (12.5,11.5) {1};
\node at (13.5,11.5) {0};
\fill[paritycolor] (14,11) rectangle (15,12);
\node at (14.5,11.5) {1};
\node at (15.5,11.5) {0};
\fill[outercolor] (0,12) rectangle (1,13);
\node at (0.5,12.5) {1};
\fill[outercolor] (1,12) rectangle (2,13);
\node at (1.5,12.5) {1};
\fill[outercolor] (2,12) rectangle (3,13);
\node at (2.5,12.5) {0};
\fill[outercolor] (3,12) rectangle (4,13);
\node at (3.5,12.5) {0};
\fill[outercolor] (4,12) rectangle (5,13);
\node at (4.5,12.5) {0};
\fill[outercolor] (5,12) rectangle (6,13);
\node at (5.5,12.5) {0};
\node at (6.5,12.5) {0};
\node at (7.5,12.5) {1};
\node at (8.5,12.5) {1};
\node at (9.5,12.5) {0};
\node at (10.5,12.5) {1};
\node at (11.5,12.5) {0};
\node at (12.5,12.5) {0};
\node at (13.5,12.5) {0};
\fill[paritycolor] (14,12) rectangle (15,13);
\node at (14.5,12.5) {0};
\node at (15.5,12.5) {1};
\fill[outercolor] (0,13) rectangle (1,14);
\node at (0.5,13.5) {1};
\fill[outercolor] (1,13) rectangle (2,14);
\node at (1.5,13.5) {1};
\fill[outercolor] (2,13) rectangle (3,14);
\node at (2.5,13.5) {1};
\fill[outercolor] (3,13) rectangle (4,14);
\node at (3.5,13.5) {1};
\fill[outercolor] (4,13) rectangle (5,14);
\node at (4.5,13.5) {1};
\fill[outercolor] (5,13) rectangle (6,14);
\node at (5.5,13.5) {1};
\node at (6.5,13.5) {1};
\node at (7.5,13.5) {1};
\node at (8.5,13.5) {0};
\node at (9.5,13.5) {1};
\node at (10.5,13.5) {0};
\node at (11.5,13.5) {1};
\node at (12.5,13.5) {0};
\node at (13.5,13.5) {1};
\fill[paritycolor] (14,13) rectangle (15,14);
\node at (14.5,13.5) {1};
\node at (15.5,13.5) {0};
\fill[outercolor] (0,14) rectangle (1,15);
\node at (0.5,14.5) {1};
\fill[outercolor] (1,14) rectangle (2,15);
\node at (1.5,14.5) {0};
\fill[outercolor] (2,14) rectangle (3,15);
\node at (2.5,14.5) {0};
\fill[outercolor] (3,14) rectangle (4,15);
\node at (3.5,14.5) {1};
\fill[outercolor] (4,14) rectangle (5,15);
\node at (4.5,14.5) {0};
\fill[outercolor] (5,14) rectangle (6,15);
\node at (5.5,14.5) {1};
\node at (6.5,14.5) {0};
\node at (7.5,14.5) {0};
\node at (8.5,14.5) {1};
\node at (9.5,14.5) {0};
\node at (10.5,14.5) {0};
\node at (11.5,14.5) {1};
\node at (12.5,14.5) {1};
\node at (13.5,14.5) {1};
\fill[paritycolor] (14,14) rectangle (15,15);
\node at (14.5,14.5) {0};
\node at (15.5,14.5) {1};
\fill[outercolor] (0,15) rectangle (1,16);
\node at (0.5,15.5) {1};
\fill[outercolor] (1,15) rectangle (2,16);
\node at (1.5,15.5) {0};
\fill[outercolor] (2,15) rectangle (3,16);
\node at (2.5,15.5) {1};
\fill[outercolor] (3,15) rectangle (4,16);
\node at (3.5,15.5) {1};
\fill[outercolor] (4,15) rectangle (5,16);
\node at (4.5,15.5) {1};
\fill[outercolor] (5,15) rectangle (6,16);
\node at (5.5,15.5) {0};
\fill[paritycolor] (6,15) rectangle (7,16);
\node at (6.5,15.5) {0};
\fill[paritycolor] (7,15) rectangle (8,16);
\node at (7.5,15.5) {0};
\fill[paritycolor] (8,15) rectangle (9,16);
\node at (8.5,15.5) {0};
\fill[paritycolor] (9,15) rectangle (10,16);
\node at (9.5,15.5) {1};
\fill[paritycolor] (10,15) rectangle (11,16);
\node at (10.5,15.5) {0};
\fill[paritycolor] (11,15) rectangle (12,16);
\node at (11.5,15.5) {1};
\fill[paritycolor] (12,15) rectangle (13,16);
\node at (12.5,15.5) {1};
\fill[paritycolor] (13,15) rectangle (14,16);
\node at (13.5,15.5) {1};
\fill[paritycolor] (14,15) rectangle (15,16);
\node at (14.5,15.5) {1};
\fill[paritycolor] (15,15) rectangle (16,16);
\node at (15.5,15.5) {1};
\draw[step=1,gray!60,thin] (0,0) grid (16,16);
\draw[thick] (0,0) rectangle (16,16);
\node at (0.5,-.48) {$ 1 $};
\node at (1.5,-.48) {$ 2 $};
\node at (2.5,-.48) {$ 3 $};
\node at (3.5,-.48) {$ 4 $};
\node at (4.5,-.48) {$ 5 $};
\node at (5.5,-.48) {$ 6 $};
\node at (6.5,-.48) {$ 7 $};
\node at (7.5,-.48) {$ 8 $};
\node at (8.5,-.48) {$ 9 $};
\node at (9.5,-.48) {$ 10 $};
\node at (10.5,-.48) {$ 11 $};
\node at (11.5,-.48) {$ 12 $};
\node at (12.5,-.48) {$ 13 $};
\node at (13.5,-.48) {$ 14 $};
\node at (14.5,-.48) {$ 15 $};
\node at (15.5,-.48) {$ 16 $};
\node at (-.5,0.5) {$ 1 $};
\node at (-.5,1.5) {$ 2 $};
\node at (-.5,2.5) {$ 3 $};
\node at (-.5,3.5) {$ 4 $};
\node at (-.5,4.5) {$ 5 $};
\node at (-.5,5.5) {$ 6 $};
\node at (-.5,6.5) {$ 7 $};
\node at (-.5,7.5) {$ 8 $};
\node at (-.5,8.5) {$ 9 $};
\node at (-.5,9.5) {$ 10 $};
\node at (-.5,10.5) {$ 11 $};
\node at (-.5,11.5) {$ 12 $};
\node at (-.5,12.5) {$ 13 $};
\node at (-.5,13.5) {$ 14 $};
\node at (-.5,14.5) {$ 15 $};
\node at (-.5,15.5) {$ 16 $};
\draw[thick] (4,0)--(4,4)--(0,4);
\draw[thick] (6,0)--(6,6)--(0,6);
\draw (0,-1)--(0,-1.2)--(4,-1.2)--(4,-1);
\node at (2.0,-1.65) {$ 4 $};
\draw (4,-1)--(4,-1.2)--(6,-1.2)--(6,-1);
\node at (5.0,-1.65) {$ 2 $};
\draw (6,-1)--(6,-1.2)--(16,-1.2)--(16,-1);
\node at (11.0,-1.65) {$ 10 $};
\draw (-1,0)--(-1.2,0)--(-1.2,4)--(-1,4);
\node at (-1.65,2.0) {$ 4 $};
\draw (-1,4)--(-1.2,4)--(-1.2,6)--(-1,6);
\node at (-1.65,5.0) {$ 2 $};
\draw (-1,6)--(-1.2,6)--(-1.2,16)--(-1,16);
\node at (-1.65,11.0) {$ 10 $};
\draw[red!75!black,line width=.8pt] (0,3) rectangle (16,4);
\draw[red!75!black,line width=.8pt] (15,0) rectangle (16,16);
\end{tikzpicture}%
}
\par\smallskip\footnotesize (c) Completed array $\bm X'$
\end{minipage}
\caption{The arrays in Example~\ref{ex:decoding}. The red outlines in (a) mark the deleted row and column. In (c), the recovered row and column are placed at positions $(i',j')=(4,16)$.}
\label{fig:example}
\end{figure}

\subsection{Encoding}
We encode $\bm S$, $\bm U$, and $\bm B$ separately, and obtain the encoders for $\bm V$ and $\bm C$ by transposition. We fill $\bm F$ and the free entries of $\bm D$ directly with message symbols. A message part of length zero is interpreted as the empty sequence. 

\begin{lemma}\label{lem:encode-S}
There is an injective encoder from $\Sigma_q^{\lfloor\log_q|\mathcal S_{q,u}|\rfloor}$ to $\mathcal S_{q,u}$. Encoding and recovering the message from the encoded block take $O_q(q^{u^2}u^2)$ time.
\end{lemma}
\begin{IEEEproof}
We first show that $\mathcal S_{q,u}$ is nonempty. For $u=3$, consider
\[
\bm S=\begin{bmatrix}
0&1&0\\
1&1&0\\
0&0&1
\end{bmatrix}.
\]
The four entries specified in~\eqref{eq:S} are zero, and the row labels are $(0,2,0)$, $(0,1,0)$, and $(1,1,0)$. Since $\bm S$ is symmetric, these are also its column labels. For $u\geq 4$, consider the binary array with entry $\ind_{\{i\geq j\}}$ at position $(i,j)$. We place its rows in the order $(2,3,\ldots,u,1)$ and its columns in the order $(2,3,\ldots,u-1,1,u)$. The first two entries of the last column and the first two entries of the last row are then zero. The numbers of zeros in the rows are pairwise distinct, as are the numbers of zeros in the columns. Since each of these numbers is the sum of the coordinates of its label, the array satisfies~\eqref{eq:S}.

We scan the $q^{u^2}$ arrays in lexicographic order, reading the entries row by row. We count only the arrays satisfying~\eqref{eq:S}. The count gives $|\mathcal S_{q,u}|$, and the message index gives the position of $\bm S$ among the valid arrays. The same scan recovers the index from $\bm S$. Testing~\eqref{eq:S} takes $O_q(u^2)$ time per array by the method in the proof of Theorem~\ref{thm:dec_c2}, so the total time is $O_q(q^{u^2}u^2)$.
\end{IEEEproof}

For $\bm U$ and $\bm B$, we use the following enumerative coding procedure~\cite{cover1973enumerative}. Let $\mathcal A_1,\ldots,\mathcal A_h$ be nonempty finite ordered sets with $M_i=|\mathcal A_i|$, and set $M=\prod_{i=1}^h M_i$. We order the tuples in $\mathcal A_1\times\cdots\times\mathcal A_h$ lexicographically and index them from zero.
Given $\bm w\in\Sigma_q^{\lfloor\log_q M\rfloor}$, we take the tuple with index $I_h=\Rank(\bm w)$. For $i=h,\ldots,1$, we divide $I_i$ by $M_i$ to obtain the quotient $I_{i-1}$ and the remainder $z_i$, so that
\[
I_i=M_i I_{i-1}+z_i,\qquad 0\le z_i<M_i.
\]
The $i$th coordinate of the tuple is the element with index $z_i$ in $\mathcal A_i$, where the indices in each factor start from zero.
Conversely, given the tuple, we read the index $z_i$ of its $i$th coordinate in $\mathcal A_i$. Starting from $I_0=0$, we apply the same recurrence for $i=1,\ldots,h$ to obtain the tuple index $I_h$. For a tuple produced by the encoder, $I_h<q^{\lfloor\log_q M\rfloor}$. We write $I_h$ using exactly $\lfloor\log_q M\rfloor$ base-$q$ digits, including leading zeros, to recover $\bm w$.

\begin{lemma}\label{lem:encode-U}
Let
\[
M_U=(q-1)^2q^{2u-4}\prod_{j=2}^{v-u-1}(q^u-j),
\]
where an empty product equals $1$. There is an injective encoder from $\Sigma_q^{\lfloor\log_q M_U\rfloor}$ to the blocks $\bm U$ satisfying~\eqref{eq:U}. Encoding and recovering the message from the encoded block take $O_q(v^2)$ time.
\end{lemma}

\begin{IEEEproof}
For $\bm w\in\Sigma_q^{\lfloor\log_q M_U\rfloor}$, take the tuple of index $\Rank(\bm w)$ in the following product, ordered lexicographically and indexed from zero,
\[
(c_1,c_2,\bm a,\bm b,z_2,\ldots,z_{v-u-1})
\in [0:q-2]^2\times\Sigma_q^{u-2}\times\Sigma_q^{u-2}
\times\prod_{j=2}^{v-u-1}[0:q^u-j-1].
\]
The product has $M_U$ elements. To obtain the tuple, we divide $\Rank(\bm w)$ successively by the numbers of choices for its coordinates, starting with the last coordinate and carrying the quotient to the next division. The remainders give the indices of the coordinates in reverse order.

We set the first and last columns of $\bm U$ to
\[
\bm U_{[1:u],1}=(c_1+1,c_2+1,\bm a)^{\mathrm T},
\qquad
\bm U_{[1:u],v-u}=(0,0,\bm b)^{\mathrm T}.
\]
These two columns are distinct and satisfy the marker conditions in~\eqref{eq:U}. We then fill the interior columns from left to right. For $j\in[2:v-u-1]$, let $E_j$ be the set of ranks in $\Sigma_q^u$ of the first $j-1$ columns and the last column of $\bm U$. There are $j$ excluded ranks, so we take the rank $t_j$ of column $j$ to be the element of index $z_j$ in the increasing list $[0:q^u-1]\setminus E_j$. The $u$ base-$q$ digits of $t_j$ give the entries of this column.

To compute $t_j$, we start with $t=z_j$ and scan $E_j$ in increasing order, increasing $t$ by one whenever the next excluded rank is at most $t$. The final value is $t_j$, since
\[
t_j\notin E_j,\qquad t_j-|\{e\in E_j:e<t_j\}|=z_j.
\]
To recover the message, we subtract one from the first two entries of the first column to obtain $c_1,c_2$, and read $\bm a,\bm b$ from the remaining entries of the first and last columns. The displayed identity gives each $z_j$ from the rank of column $j$. The index of the recovered tuple is $\Rank(\bm w)$, which gives the message as described above.

There are $O(v)$ columns, and maintaining and scanning the ordered list of excluded ranks takes $O(v)$ time per column. The number of choices for each coordinate of the tuple is at most $q^u$, which fits in one word. Since $M_U\le q^{u(v-u)}$ and $v=O_q(\log n)$, both $M_U$ and the tuple index occupy $O_q(u)$ words. Each multiplication or division by a factor size takes $O_q(u)$ time. There are $O(v)$ factors, so computing $M_U$ and converting between the tuple and its index take $O_q(uv)$ time. Converting the message to or from its index also takes $O_q(uv)$ time. Thus encoding and message recovery both take $O_q(v^2)$ time.
\end{IEEEproof}

\begin{lemma}\label{lem:encode-B}
Let $\bm a\in\Sigma_q^v$, let $\sigma$ be a permutation of $[1:v]$, and let $\beta\in[0:n-v-1]$. Set
\[
M_B=\bigl((q-1)^2q^{v-2}-1\bigr)(q^v-2)^{n-v-5}q^{2v-6}.
\]
There is an injective encoder from $\Sigma_q^{\lfloor\log_q M_B\rfloor}$ to the blocks $\bm B\in\Sigma_q^{v\times(n-v)}$ with the following properties.
\begin{itemize}
    \item The first two entries of the first column are nonzero, and any two adjacent columns are distinct.
    \item The sum of the columns is $\bm a$ modulo $q$.
    \item $\VTSyn(\Rank^c(\widehat{\bm B}))=\beta$, where $\widehat B_{i,j}=B_{\sigma(i),j}$.
\end{itemize}
Encoding and recovering the message from the encoded block take $O_q(n^2)$ time.
\end{lemma}

\begin{IEEEproof}
Let $\bm m=(q-1,\ldots,q-1)^{\mathrm T}\in\Sigma_q^v$ be the column of length $v$ whose entries are all $q-1$. We first construct $\bm B^-\in\Sigma_q^{v\times(n-v-1)}$ with distinct adjacent columns, column sum $\bm a-\bm m$ modulo $q$, and no occurrence of $\bm m$. We then insert $\bm m$ into $\bm B^-$ to obtain the prescribed column sum and VT syndrome.

For $\bm w\in\Sigma_q^{\lfloor\log_q M_B\rfloor}$, take the tuple of index $\Rank(\bm w)$ in the following product, ordered lexicographically with indices starting from zero,
\[
\begin{aligned}
&(z_1,\ldots,z_{n-v-4},\bm e_{[1:v-3]},\bm f_{[1:v-3]}) \in [0:(q-1)^2q^{v-2}-2]\times[0:q^v-3]^{n-v-5}
\times\Sigma_q^{v-3}\times\Sigma_q^{v-3}.
\end{aligned}
\]
This product has $M_B$ elements. We obtain the tuple by successive divisions as in the proof of Lemma~\ref{lem:encode-U}. The indices $z_1,\ldots,z_{n-v-4}$ specify the first $n-v-4$ columns of $\bm B^-$. The last two coordinates specify the free entries of two of its last three columns.

We first fill the columns $\bm d_1,\ldots,\bm d_{n-v-4}\in\Sigma_q^v$ from left to right. To obtain $\bm d_1$, we take the tuple of index $z_1$ in $[0:q-2]^2\times\Sigma_q^{v-2}$ and add one to its first two entries. The resulting column has nonzero first two entries. The last tuple in this product would give $\bm m$, and the range of $z_1$ excludes that tuple. For $j\in[2:n-v-4]$, we take $\bm d_j$ to be the column of index $z_j$ in the lexicographic list obtained by excluding $\bm d_{j-1}$ and $\bm m$ from $\Sigma_q^v$. Since $\bm m$ has the largest rank in $\Sigma_q^v$, the rank of $\bm d_j$ is
\[
\Rank(\bm d_j)=z_j+\ind_{\{z_j\geq\Rank(\bm d_{j-1})\}}.
\]
The $v$ base-$q$ digits of this rank give the entries of $\bm d_j$.

We next fill the last three columns $\bm e,\bm f,\bm g\in\Sigma_q^v$. The tuple gives the first $v-3$ entries of $\bm e$ and $\bm f$. Let $\bm h$ be the reduction of $\bm a-\sum_{j=1}^{n-v-4}\bm d_j$ modulo $q$. We need the last three columns to sum to $\bm h-\bm m$ modulo $q$. To ensure that neighboring columns differ and that none of these columns equals $\bm m$, we set the last three entries of $\bm e$ and $\bm f$ as follows, with all arithmetic modulo $q$.
\[
\begin{array}{c|cc}
\text{Coordinate}&\bm e&\bm f\\ \hline
v-2&0&1\\
v-1&(\bm d_{n-v-4})_{v-1}+1&h_{v-1}-e_{v-1}+1\\
v&h_v&0
\end{array}
\]
We then set $\bm g=\bm h-\bm e-\bm f-\bm m$ modulo $q$. Since $m_{v-1}=m_v=q-1$, the prescribed entries give $g_{v-1}=0$ and $g_v=1$. Thus $\bm d_{n-v-4}$ and $\bm e$ differ at coordinate $v-1$, $\bm e$ and $\bm f$ differ at coordinate $v-2$, and $\bm f$ and $\bm g$ differ at coordinate $v$. Moreover, $e_{v-2}=f_v=g_{v-1}=0$, so each of these three columns differs from $\bm m$. Therefore
\[
\bm B^-=[\bm d_1\ \cdots\ \bm d_{n-v-4}\ \bm e\ \bm f\ \bm g]
\]
has the required column sum, has distinct adjacent columns, and omits $\bm m$.

We read each column of $\bm B^-$ in row order $\sigma$ and take its rank in $\Sigma_q^v$. In this order, $\bm m$ still has rank $q^v-1$. Since this rank is absent from the sequence for $\bm B^-$, its $n-v$ insertion positions give $n-v$ distinct sequences. All these sequences have the same symbol sum modulo $q^v$, and deleting the inserted rank from any of them gives the same subsequence. By Lemma~\ref{thm:vt}, their VT syndromes are pairwise distinct. Since the syndromes are taken modulo $n-v$, exactly one insertion position gives $\beta$. We apply Corollary~\ref{cor:known-symbol} with the rank sequence of $\bm B^-$, the known rank $q^v-1$, and the syndrome $\beta$ to find this position. Inserting $\bm m$ there gives $\bm B$ with column sum $\bm a$ modulo $q$. Its adjacent columns are distinct because $\bm m$ differs from every column of $\bm B^-$. Its first column is either $\bm d_1$ or $\bm m$, both of which have nonzero first two entries.

To recover the message from $\bm B$, we delete its unique occurrence of $\bm m$ to obtain $\bm B^-$. We recover $z_1$ by subtracting one from the first two entries of $\bm d_1$ and taking the index of the resulting tuple in $[0:q-2]^2\times\Sigma_q^{v-2}$. For $j\in[2:n-v-4]$, the column $\bm d_j$ gives
\[
z_j=\Rank(\bm d_j)
-\ind_{\{\Rank(\bm d_j)>\Rank(\bm d_{j-1})\}}.
\]
The first $v-3$ entries of $\bm e$ and $\bm f$ give the last two coordinates of the tuple. We then compute the index of the recovered tuple in lexicographic order and write it using $\lfloor\log_q M_B\rfloor$ base-$q$ digits to recover $\bm w$.

There are $O(n)$ columns, and constructing or reading each column takes $O_q(v)$ time, including the computation of its rank and its contribution to the column sum. The number of choices for each coordinate of the tuple is at most $q^v=O_q(n)$, which fits in one word. Since $M_B\leq q^{v(n-v-1)}$ and $v=O_q(\log n)$, both $M_B$ and the tuple index occupy $O_q(n)$ words. Each multiplication or division by a factor size takes $O_q(n)$ time. There are $O(n)$ factors, so computing $M_B$ and converting between the tuple and its index take $O_q(n^2)$ time. Converting the message to or from its index takes $O_q(v(n-v))$ time. Once the column ranks are available, finding the VT insertion position takes $O(n-v)$ time. Thus encoding and message recovery both take $O_q(n^2)$ time.
\end{IEEEproof}

Let the message consist of seven parts $\bm w_i\in\Sigma_q^{l_i}$, corresponding to $\bm S$, $\bm U$, $\bm V$, $\bm F$, $\bm B$, $\bm C$, and $\bm D$, respectively. The lengths of these parts are
\begin{equation}\label{eq:message-lengths}
\begin{alignedat}{2}
l_1&=\lfloor\log_q|\mathcal S_{q,u}|\rfloor,
&\qquad l_2=l_3&=\lfloor\log_q M_U\rfloor,\\
l_4&=(v-u)^2,
& l_5=l_6&=\lfloor\log_q M_B\rfloor,\\
l_7&=(n-v-1)^2. &&
\end{alignedat}
\end{equation}

\begin{theorem}\label{thm:enc_c2}
For $q\geq 2, n\geq q^4, u=\lceil\log_q(\lceil\log_qn\rceil^2)\rceil\geq 3, v=\lceil\log_qn\rceil+2$ and $l_i$ defined in~\eqref{eq:message-lengths}, there is an injective encoder from $\prod_{i=1}^7\Sigma_q^{l_i}$ to $\mathcal C^2_{q,n}$. Encoding and recovering the message from the encoded array take $O_q(n^2)$ time. In particular, the image $\mathcal C^2_{\mathrm{enc}}$ has size $q^{\sum_{i=1}^7 l_i}$.
\end{theorem}
\begin{IEEEproof}
We first apply Lemma~\ref{lem:encode-S} to $\bm w_1$ to obtain $\bm S$. By Lemma~\ref{lem:encode-U}, we encode $\bm w_2$ into $\bm U$ and $\bm w_3$ into $\bm V^{\mathrm T}$. We fill $\bm F$ with $\bm w_4$ in row order. These four blocks give $\bm A$ and determine $\sigma_r$ and $\sigma_c$.

We next construct $\bm B$ and $\bm C$ using Lemma~\ref{lem:encode-B}. By~\eqref{eq:B} and~\eqref{eq:C}, the prescribed sums of the columns of $\bm B$ and $\bm C^{\mathrm T}$ are, respectively,
\[
\bm a_c=-\sum_{j=1}^v\bm A_{[1:v],j}\pmod q,
\qquad
\bm a_r=-\left(\sum_{i=1}^v\bm A_{i,[1:v]}\right)^{\mathrm T}\pmod q.
\]
We encode $\bm w_5$ into $\bm B$ with sum $\bm a_c$, row order $\sigma_r$, and syndrome $\VTSyn(\sigma_c^{-1})$. Similarly, we encode $\bm w_6$ into $\bm C^{\mathrm T}$ with sum $\bm a_r$, row order $\sigma_c$, and syndrome $\VTSyn(\sigma_r^{-1})$. Both syndromes lie in $[0:v-1]\subseteq[0:n-v-1]$. Reading $\bm C^{\mathrm T}$ in row order $\sigma_c$ gives $(\widehat{\bm C})^{\mathrm T}$, so the constraint on the column ranks of this transposed block gives the row constraint in~\eqref{eq:C}.

Finally, we fill the upper-left $(n-v-1)\times(n-v-1)$ subarray of $\bm D$ with $\bm w_7$ in row order. We use the row sums in~\eqref{eq:D} to fill the last entry of each of the first $n-v-1$ rows. We then use the column sums to fill the last row. Since the prescribed row sums and column sums have the same total modulo $q$, the last row also has the required sum. This gives an array in $\mathcal C^2_{q,n}$.

Lemmas~\ref{lem:encode-S}--\ref{lem:encode-B} recover the message parts encoded in $\bm S$, $\bm U$, $\bm V^{\mathrm T}$, $\bm B$, and $\bm C^{\mathrm T}$. We read $\bm w_4$ directly from $\bm F$ and $\bm w_7$ from the upper-left subarray of $\bm D$. Thus the encoder is injective.

By Lemma~\ref{lem:encode-S}, encoding $\bm S$ and recovering its message take $O_q(q^{u^2}u^2)$ time. Since $u=O_q(\log\log n)$, this is $n^{o(1)}$ for fixed $q$. By Lemma~\ref{lem:encode-U}, the corresponding operations for $\bm U$ and $\bm V$ each take $O_q(v^2)$ time. Filling $\bm F$ and computing the canonical orders and the prescribed sums from $\bm A$ also take $O_q(v^2)$ time.
By Lemma~\ref{lem:encode-B}, encoding and message recovery for $\bm B$ and $\bm C$ each take $O_q(n^2)$ time. Filling $\bm D$ and reading its message symbols each take $O_q(n^2)$ time. Thus encoding the array and recovering the message from it each take $O_q(n^2)$ time.
\end{IEEEproof}

\subsection{Redundancy}
To bound the redundancy of $\mathcal C^2_{\mathrm{enc}}$, we estimate the number of choices for each message field. For $\bm U$ and $\bm V$, we choose $v-u$ distinct vectors from an alphabet of size $q^u\geq (v-u)^2$. For $\bm B$ and $\bm C$, each of the intermediate columns has only two forbidden values out of $q^v\geq q^2n$. The corresponding products therefore cost a bounded number of symbols. It remains to bound the cost of the distinct labels in $\bm S$.

\begin{lemma}\label{lem:Ssize}
Let $q\geq 2$ and set $H_q=3+\left\lceil\frac{q^2}{q-1}\left(3+\ln\frac{q}{2}\right)^2\right\rceil$. If $u\geq H_q$, then the family $\mathcal S_{q,u}$ defined by~\eqref{eq:S} satisfies
\[
|\mathcal S_{q,u}|>\frac{q^{u^2-4}}{32}.
\]
\end{lemma}
\begin{IEEEproof}
Choose $\bm S$ uniformly subject to $S_{u,1}=S_{u,2}=S_{1,u}=S_{2,u}=0$, and fix two distinct rows $i_1,i_2$. The $u-3$ columns outside $\{1,2,u\}$ are independent and uniform. For $d\in[1:u]$, set
\[
p_d=\frac{2(q-1)}{q^2}\Pr\left(\mathrm B\left(u-2,\frac1q\right)=d-1\right).
\]
Each of these columns contributes $\bm e_d$ or $-\bm e_d$ to $\lambda_{i_1}(\bm S)-\lambda_{i_2}(\bm S)$ with probability $p_d/2$ each, where $\bm e_d$ is the $d$th standard basis vector of $\mathbb Z^u$. With the remaining probability, the contribution is zero. Let $\bm Z$ be the sum of these contributions, and let $\bm T$ be the contribution of columns $1$, $2$, and $u$. Since $\bm Z$ and $\bm T$ are independent, we have
\[
\begin{aligned}
\Pr\bigl(\lambda_{i_1}(\bm S)=\lambda_{i_2}(\bm S)\bigr)
&=\sum_{\bm t\in\mathbb Z^u}\Pr(\bm T=-\bm t)\Pr(\bm Z=\bm t)
\leq\sup_{\bm z\in\mathbb Z^u}\Pr(\bm Z=\bm z).
\end{aligned}
\]

We first bound the largest point probability of $\bm Z$. Since $\sum_{d=1}^u p_d=2(q-1)/q^2\leq 1/2$, we can generate each column contribution as follows. We select coordinate $d$ with probability $2p_d$ and place the sum of two independent fair Bernoulli variables minus one in that coordinate, with all other coordinates zero. With the remaining probability, we return the zero vector. The coordinate selections are independent across columns, and all Bernoulli variables are independent of these selections.

Let $K_d$ be the number of times coordinate $d$ is selected. Conditional on $(K_1,\ldots,K_u)$, the coordinates of $\bm Z$ are independent, and $Z_d+K_d$ has distribution $\mathrm B(2K_d,1/2)$. Each conditional point probability is therefore largest at $Z_d=0$. Consequently,
\[
\sup_{\bm z\in\mathbb Z^u}\Pr(\bm Z=\bm z)
=\Pr(\bm Z=\bm0)
=\E\left[\prod_{d=1}^u4^{-K_d}\binom{2K_d}{K_d}\right].
\]

Integration by parts gives $4^{-k}\binom{2k}{k}=\frac2\pi\int_0^{\pi/2}\cos^{2k}t\,dt$ for $k\geq 0$. Using this identity and the multinomial theorem, we obtain
\[
\begin{aligned}
\Pr(\bm Z=\bm0)
&=\left(\frac2\pi\right)^u\int_{[0,\pi/2]^u}
\left(1-2\sum_{d=1}^u p_d\sin^2t_d\right)^{u-3}\,d\bm t
\leq\prod_{d=1}^u\frac2\pi\int_0^{\pi/2}
e^{-2(u-3)p_d\sin^2t}\,dt.
\end{aligned}
\]
Here the quantity raised to the power $u-3$ is nonnegative because $2\sum_{d=1}^u p_d\leq 1$, and the inequality follows from $1-z\leq e^{-z}$. For $x\geq 0$, applying $e^{-2z}\leq(1+z)^{-2}$ for $z\geq 0$ and then substituting $s=\tan t$ gives
\[
\frac2\pi\int_0^{\pi/2}e^{-2x\sin^2t}\,dt
\leq\frac2\pi\int_0^{\pi/2}\frac{dt}{(1+x\sin^2t)^2}
=\frac{1+x/2}{(1+x)^{3/2}}.
\]

Set $f(x)=\frac32\ln(1+x)-\ln(1+x/2)$ and $y=\sqrt{(q-1)(u-3)}/q$. The function $f$ is increasing and concave on $[0,\infty)$, with $f(0)=0$. By Lemma~\ref{lem:bin} and the definition of $p_d$,
\[
(u-3)\max_d p_d
\leq\sqrt{\frac2e}\sqrt{\frac{u-3}{u-2}}\,y
<\frac9{10}y,
\qquad
(u-3)\sum_{d=1}^u p_d=2y^2.
\]
By concavity, $f(x)\geq x f(9y/10)/(9y/10)$ for $0\leq x\leq9y/10$. Combining the preceding bounds, we obtain
\[
\begin{aligned}
\Pr\bigl(\lambda_{i_1}(\bm S)=\lambda_{i_2}(\bm S)\bigr)
&\leq\Pr(\bm Z=\bm0)
\leq\exp\left(-\sum_{d=1}^u f\bigl((u-3)p_d\bigr)\right)
\leq\exp\left(-\frac{20}{9}y f\left(\frac{9y}{10}\right)\right).
\end{aligned}
\]
The four prescribed entries are invariant under transposition, so the same bound holds for any pair of column labels.

Let $E$ be the event that two row labels are equal or two column labels are equal. Taking the union bound over the $2\binom u2<u^2$ pairs gives
\begin{equation}\label{eq:seed-density-tail}
\Pr(E)\leq u^2\exp\left(-\frac{20}{9}y f\left(\frac{9y}{10}\right)\right).
\end{equation}
Since $u\geq H_q$, we have $y\geq3+\ln(q/2)\geq3$. Thus
\[
u=3+\frac{q^2y^2}{q-1}
\leq3+2qy^2
\leq3+4e^{y-3}y^2.
\]
It follows from~\eqref{eq:seed-density-tail} that $\Pr(E)\leq F(y)$, where
\[
F(y)=\left(3+4e^{y-3}y^2\right)^2
\exp\left(-\frac{20}{9}y f\left(\frac{9y}{10}\right)\right).
\]

We now show that $F$ is decreasing on $[3,\infty)$. For $x\geq27/10$, we have $f(x)\geq f(27/10)>1$, since $(37/10)^3/(47/20)^2>9>e^2$. Also,
\[
xf'(x)=\frac12+\frac{x-2}{2(x+1)(x+2)}>\frac12.
\]
Therefore, for $y\geq3$,
\[
\begin{aligned}
\frac{d}{dy}\ln F(y)
&=\frac{8e^{y-3}y^2}{3+4e^{y-3}y^2}\left(1+\frac2y\right) -\frac{20}{9}\left[f\left(\frac{9y}{10}\right)
+\frac{9y}{10}f'\left(\frac{9y}{10}\right)\right]
<2+\frac43-\frac{20}{9}\left(1+\frac12\right)=0.
\end{aligned}
\]
Hence
\[
\Pr(E)\leq F(y)\leq F(3)
=39^2\frac{(47/20)^{20/3}}{(37/10)^{10}}
<\frac{31}{32}.
\]
Thus more than $1/32$ of the $q^{u^2-4}$ arrays satisfying the four prescribed entries have pairwise distinct row labels and pairwise distinct column labels. These arrays belong to $\mathcal S_{q,u}$, which proves the result.
\end{IEEEproof}

\begin{theorem}\label{thm:encoded-redundancy}
Let $q\geq 2$ and $n\geq q^4$. Set
\[
u=\left\lceil\log_q\bigl(\lceil\log_q n\rceil^2\bigr)\right\rceil,
\qquad
v=\lceil\log_q n\rceil+2,
\]
and assume $u\geq 3$. Let $H_q=3+\left\lceil\frac{q^2}{q-1}\left(3+\ln\frac{q}{2}\right)^2\right\rceil$. Then the image $\mathcal C^2_{\mathrm{enc}}$ of the encoder in Theorem~\ref{thm:enc_c2} satisfies
\[
r_{\mathcal C^2_{\mathrm{enc}}}\leq 2n+2\lceil\log_q n\rceil+31+(H_q-1)^2.
\]
If $u\geq H_q$, then
\[
r_{\mathcal C^2_{\mathrm{enc}}}\leq 2n+2\lceil\log_q n\rceil+40.
\]
In particular, for each fixed $q\geq 2$ and all sufficiently large $n$, the redundancy is $2n+2\log_q n+O_q(1)$.
\end{theorem}
\begin{IEEEproof}
By Theorem~\ref{thm:enc_c2}, we have $r_{\mathcal C^2_{\mathrm{enc}}}=n^2-\sum_{i=1}^7l_i$. We bound the contribution of each block to the redundancy by subtracting its message length from its number of entries.

For $\bm S$, the nonemptiness of $\mathcal S_{q,u}$ gives $l_1\geq 0$. Thus, if $u<H_q$, then $u^2-l_1\leq (H_q-1)^2$. If $u\geq H_q$, Lemma~\ref{lem:Ssize} gives $u^2-l_1\leq 9$.

For $\bm U$ and $\bm V$, the expression for $M_U$ in Lemma~\ref{lem:encode-U} gives
\[
\frac{M_U}{q^{u(v-u)}}=\frac{(q-1)^2}{q^4}\prod_{j=2}^{v-u-1}\left(1-\frac{j}{q^u}\right).
\]
Since $q^u\geq (v-u)^2$, we have $\sum_{j=2}^{v-u-1}j/q^u<1/2$. Applying $\prod_j(1-x_j)\geq 1-\sum_jx_j$ for $0\leq x_j\leq 1$, we obtain
\[
\frac{M_U}{q^{u(v-u)}}>\frac{(q-1)^2}{2q^4}\geq q^{-5},
\]
where the last inequality follows from $q(q-1)^2\geq 2$. Since $l_2=l_3=\lfloor\log_q M_U\rfloor$, it follows that $u(v-u)-l_2=u(v-u)-l_3\leq 5$.

For $\bm B$ and $\bm C$, the expression for $M_B$ in Lemma~\ref{lem:encode-B} gives
\[
\frac{M_B}{q^{v(n-v)}}=q^{-2v-6}\left(\frac{(q-1)^2}{q^2}-q^{-v}\right)(1-2q^{-v})^{n-v-5}.
\]
Since $q^v\geq q^2n\geq 64$, the middle factor is at least $15/64$. Also, using $-\ln(1-x)\leq x/(1-x)$ for $0<x<1$ and $n-v-5\leq n$, we have
\[
-(n-v-5)\log_q(1-2q^{-v})
\leq\frac{2q^{-2}}{(1-2q^{-v})\ln q}
\leq\frac{16}{31\ln 2}.
\]
Therefore,
\[
v(n-v)-\log_q M_B
\leq 2v+6+\log_2\frac{64}{15}+\frac{16}{31\ln 2}
<2v+9.
\]
Taking the ceiling and using $l_5=l_6=\lfloor\log_q M_B\rfloor$, we obtain $v(n-v)-l_5=v(n-v)-l_6\leq 2v+9$.

The block $\bm F$ contributes no redundancy, while $\bm D$ contributes $(n-v)^2-(n-v-1)^2=2(n-v)-1$. Adding the contributions gives
\[
\begin{aligned}
r_{\mathcal C^2_{\mathrm{enc}}}
&\leq (u^2-l_1)+2\cdot 5+2(2v+9)+2(n-v)-1\\
&=(u^2-l_1)+2n+2v+27.
\end{aligned}
\]
Since $9\leq (H_q-1)^2$, the bound $u^2-l_1\leq (H_q-1)^2$ holds in both cases. Substituting $v=\lceil\log_q n\rceil+2$ gives the first bound. If $u\geq H_q$, using $u^2-l_1\leq 9$ gives the second bound. For fixed $q$, we have $u\to\infty$ as $n\to\infty$, so the second bound applies for all sufficiently large $n$. Together with Corollary~\ref{cor:lowerbound}, this gives the stated asymptotic redundancy.
\end{IEEEproof}

\section{Conclusion}\label{sec:conclusion}
For each fixed $q$, we establish a sphere-packing lower bound on the redundancy of $(1,1)$-criss-cross deletion-correcting codes. Our first construction matches this bound up to a vanishing term, achieving redundancy $2n-1+2\log_q n+o_q(1)$ with an $O_q(n^2)$-time decoder. Our second construction achieves redundancy $2n+2\log_q n+O_q(1)$ and admits $O_q(n^2)$-time algorithms for encoding, deletion correction, and message recovery.
For general fixed $t_r$ and $t_c$, it remains open to determine the constant term in the minimum redundancy for arbitrary row and column deletions and to construct efficiently encodable codes within an additive constant of the sphere-packing lower bound in~\cite{sun2026}.

\end{document}